\documentclass[aps,twocolumn,prl,superscriptaddress,amsmath,showpacs,tightenlines]{revtex4-1}
\usepackage{amssymb}
\usepackage{braket}
\usepackage{amsmath}
\usepackage{graphicx}
\usepackage{subfigure}
\usepackage{natbib}
\usepackage{epsfig}
\usepackage{amsfonts}
\usepackage{mathrsfs}
\usepackage{CJK}
\usepackage{xcolor}
\usepackage[toc,page,title,titletoc,header]{appendix}
\begin{document}
\title{Controllable superradiance scaling in photonic waveguide}
\author{Xiang Guo}
\author{Zhihai Wang}
\email{wangzh761@nenu.edu.cn}
\affiliation{Center for Quantum Sciences and School of Physics, Northeast Normal University, Changchun 130024, China}
\begin{abstract}
We investigate the superradiance of two-level target atoms (TAs) coupled to a photonic waveguide, demonstrating that the scaling of the superradiance strength can be controlled on demand by an ensemble of control atoms (CAs). The scaling with respect to the number of TAs can be lower, higher, or equal to the traditional Dicke superradiance, depending on the relative positioning of the ensembles and the type of CAs (e.g., small or giant). These phenomena are attributed to unconventional atomic correlations. Furthermore, we observe chiral superradiance of the TAs, where the degree of chirality can be enhanced by giant CAs instead of small ones. The effects discussed in this work could be observed in waveguide QED experiments, offering a potential avenue for manipulating superradiance.
\end{abstract}
\maketitle
\emph{Introduction}-The scaling behavior plays a crucial role in understanding the universality of quantum systems and exploring related applications in quantum technologies. A notable example is the Dicke superradiance phenomenon~\cite{dicke1954,MG1982}, which arises from collective light-matter interactions~\cite{MD2003,Id2008,KH2010}. Dicke superradiance demonstrates how \( N \) closely spaced emitters exhibit enhanced collective emission, with the emission strength scaling as \( N^2 \) due to cooperative interactions. This phenomenon has inspired extensive research in superradiant lasers~\cite{FH1993,JG2012,HL2020,SH2024}, driven-dissipative phase transitions~\cite{WK2013,XW2016,JH2018,MS2020,FB2024,GL2024}, and quantum precision measurements~\cite{MA2016,MA2018,VP2019,YZ2022,MK2022,HY2024}.

Recently, superradiance has garnered significant attention in waveguide QED systems~\cite{FD2019,wang2020,FD2020,SC2023,AS2023}, where emitters such as atoms~\cite{AG2015,YZ2017,RA2022,RP2022}, quantum dots~\cite{VI2010,PT2016,JH2018x,AG2020,CZ2024}, and superconducting qubits~\cite{JA2014,NL2016,JD2021,EK2021} couple to a waveguide. The waveguide acts as both a structured environment and an effective data bus, mediating interactions between emitters. Despite extensive studies, how can the scaling of the superradiance strength be controlled as the number of participating emitters changes remains underexplored.

The mean-field approximation effectively addresses superradiance in the thermodynamic limit, while the master equation captures quantum correlations. However, the master equation for the atom-photon hybrid system becomes computationally prohibitive as the number of atoms and photons increases. Therefore, developing methods to study the scaling of superradiance remains a key challenge.

In this Letter, we investigate the superradiance of target atoms (TAs) in a one-dimensional photonic waveguide, demonstrating how its scaling can be controlled by control atoms (CAs). Using the discrete truncated Wigner approximation (DTWA)~\cite{JS2015x,RK2020,LH2021,VP2022,JH2022,DD2024}, we analyze the atomic and photonic dynamics, even in the presence of waveguide dissipation. When the number of CAs (\( N_C \)) is comparable to that of the TAs (\( N_T \)), the scaling of superradiance can be tuned to exceed or fall below the standard Dicke \( N_T^2 \) law, for both of small and giant CAs ~\cite{MV2014,AF2014,LG2020,WZ2020,XW2021,AM2021,AF2018,BK2020,ZQ2022}. When \( N_T \gg N_C \), the CA control weakens, and the radiance strength reverts to the \( N_T^2 \) scaling of Dicke superradiance. In the opposite limit of \( N_C \gg N_T \), the radiance strength naturally maintains the \( N_T^2 \) scaling but gradually becomes independent of \( N_C \) as \( N_C \) increases. These scaling behaviors are further explained by the atomic correlation of TAs.

The DTWA approach also allows us to track the dynamics of the emitted photons in the waveguide. We observe interference-induced chiral superradiance, where the degree of chirality can be enhanced in the giant CAs setup compare with small ones. Furthermore, perfect chirality can be realized under the condition of \( N_C \gg N_T \).

\begin{figure}
  \includegraphics[width=1\columnwidth]{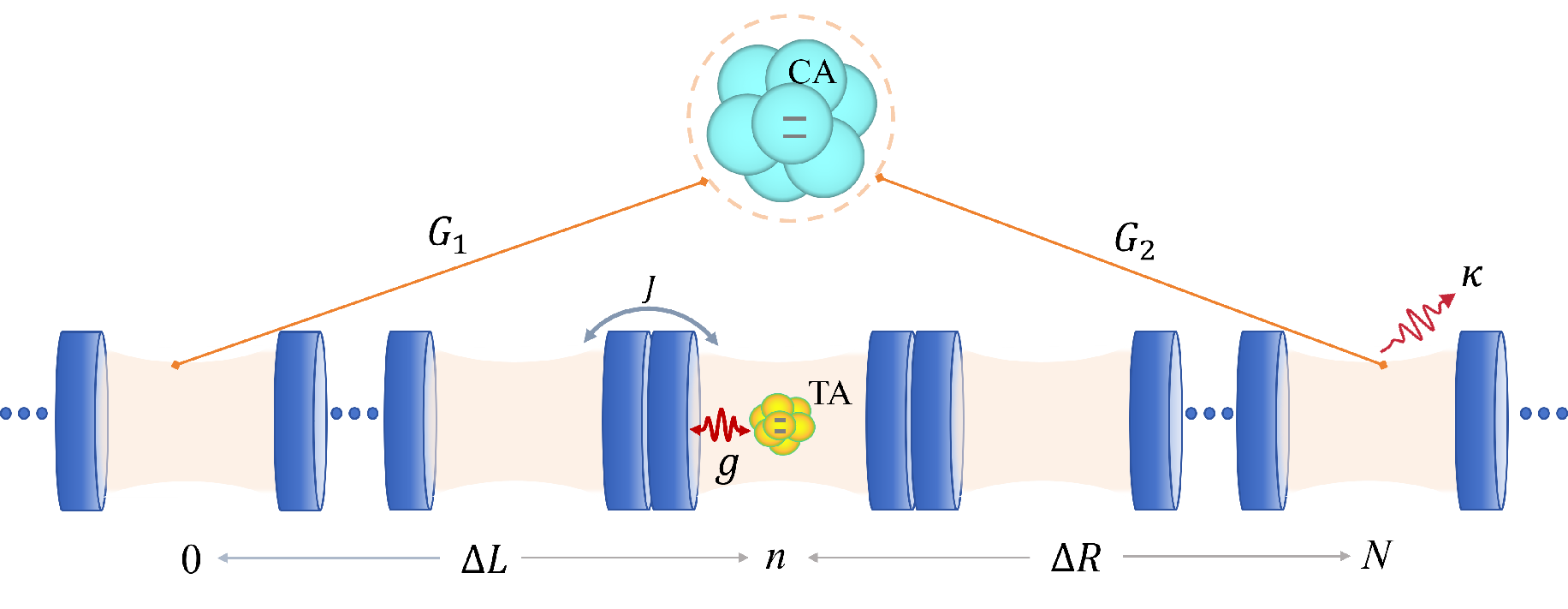}
  \caption{A sketch of the superradiance of target atoms (TA) in the photonic waveguide, manipulated by the control atoms (CA). The photonic waveguide consists of coupled resonators with nearest-neighbor tunneling rate $J$ and decay rate $\kappa$. }
  \label{scheme}
\end{figure}

\emph{Model}-As shown in Fig.~\ref{scheme}, we consider two ensembles of two-level atoms, with ground state \(|g\rangle\) and excited state \(|e\rangle\), interacting with a one-dimensional coupled resonator waveguide (CRW). The ensemble in the \(n\)th resonator is called the TA, while the ensemble coupled to the waveguide at the \(0\)th and \(N\)th sites is referred to as the CA. Both TAs and CAs can be realized with superconducting qubits, such as transmons~\cite{NL2016,JD2021}.

The CRW is modeled by the tight-binding Hamiltonian (\(\hbar = 1\))
\begin{equation}
\hat{H}_c = \sum_{j}\omega \hat{a}_{j}^{\dagger} \hat{a}_{j} - J(\hat{a}_{j+1}^\dagger \hat{a}_{j} + \hat{a}_{j}^\dagger \hat{a}_{j+1}),
\end{equation}
where \(\hat{a}_j\) is the bosonic annihilation operator for the \(j\)-th site, and the waveguide supports a continuous band centered at \(\omega\) with width \(4J\).

The full system Hamiltonian is
\begin{eqnarray}
\hat{H}&=& \hat{H}_c + \frac{\omega_{T}}{2}\sum_{i=1}^{N_T}\hat{\sigma}_{T,z}^{(i)} + \frac{\omega_{C}}{2}\sum_{i=1}^{N_{C}}\hat{\sigma}_{C,z}^{(i)}\nonumber \\
&&+\left[ g\sum_{i=1}^{N_T} \hat{a}_{n}^{\dagger} \hat{\sigma}_{T,-}^{(i)} + \sum_{i=1}^{N_{C}}\left(G_1 \hat{a}_{0}^{\dagger} + G_2 \hat{a}_{N}^{\dagger}\right) \hat{\sigma}_{C,-}^{(i)} + \mathrm{H.C.} \right]\nonumber \\
\label{Hamil}
\end{eqnarray}
where \(\omega_{T(C)}\) are the transition frequencies for TAs (CAs), and \(\hat{\sigma}_{T(C),o}^{(i)}\) (\(o = z, \pm\)) are the Pauli operators for the $i$th TA (CA). \(g\) is the coupling strength between TAs and CRW, and \(G_1, G_2\) are the coupling strengths of the CAs to the CRW. The CAs are ``giant atoms"~\cite{MV2014,AF2014,LG2020,WZ2020,XW2021,AM2021,AF2018,BK2020,ZQ2022} when \(G_1 \neq 0\) and \(G_2 \neq 0\), leading to nonlocal coupling, interference, and retardation effects.

We also consider that each resonator is immersed in an individual dissipation channel. The effect of the environment on the atom-waveguide system is described by the master equation
\begin{equation}
\frac{d}{dt}\hat{\rho} = -i[\hat{H}, \hat{\rho}] + \kappa \sum_j \left( 2 \hat{a}_j \hat{\rho} \hat{a}_j^\dagger - \hat{\rho} \hat{a}_j^\dagger \hat{a}_j - \hat{a}_j^\dagger \hat{a}_j \hat{\rho} \right),
\label{master}
\end{equation}
where \(2\kappa\) is the photon decay rate for each resonator, and spontaneous emission from the atoms is neglected. The length of the CRW, \(N_W\), is much larger than the size of CAs, \(N_W \gg N\). The Liouville space's high dimensionality, $d = (N_C+1)\times(N_T+1)\times N_0^{N_W}$ (with \(N_0\) being the photon number cutoff in each resonator), even by considering the atomic exchange symmetry, makes solving this master equation numerically infeasible. To address this, we use the DTWA technique to investigate both atomic and photonic dynamics during the superradiance of TAs within the engineered CAs. The DTWA approximates the system by transforming operators into average values and introduces Monte Carlo samples~\cite{JS2015x} for initial conditions of the classical equations. We further consider Gaussian noise at each evolution step to capture quantum fluctuations. Then, the final results are obtained by averaging the multiple trajectories. For the detailed introduction and calculations about DTWA, we refer to the supplementary materials (SM)~\cite{SM}.

\begin{figure}
  \includegraphics[width=0.48\columnwidth]{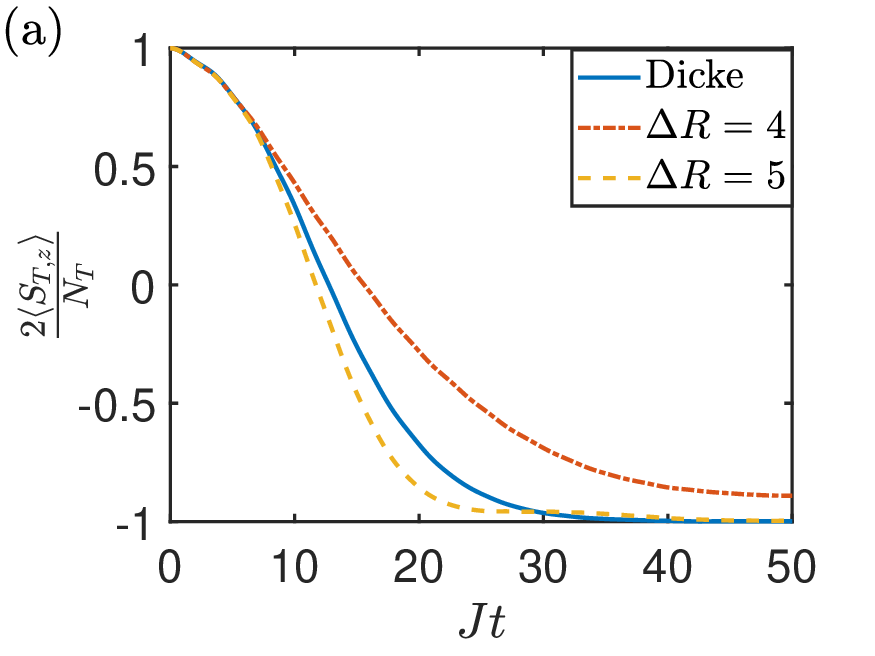}
  \includegraphics[width=0.48\columnwidth]{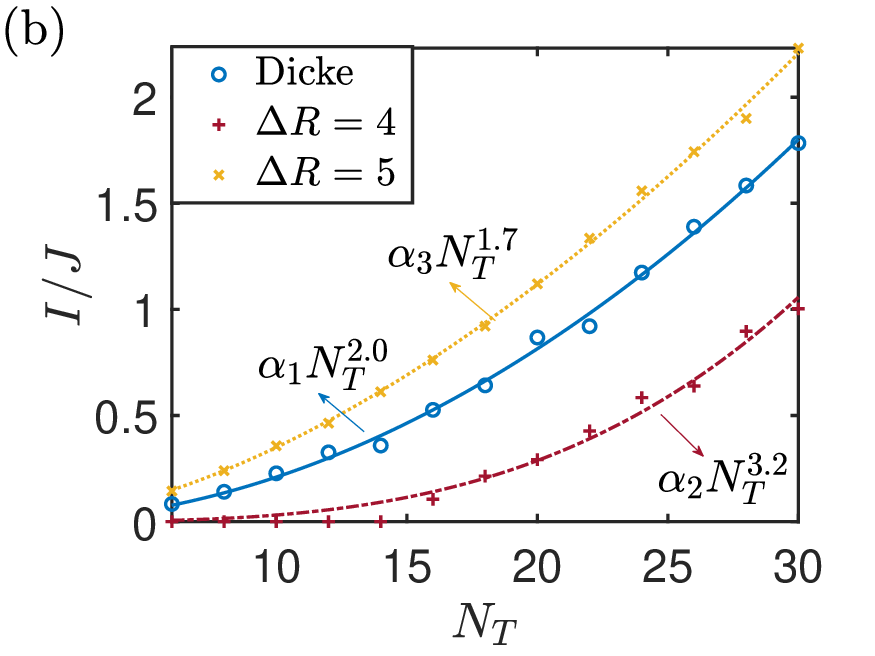}\\
  \includegraphics[width=0.48\columnwidth]{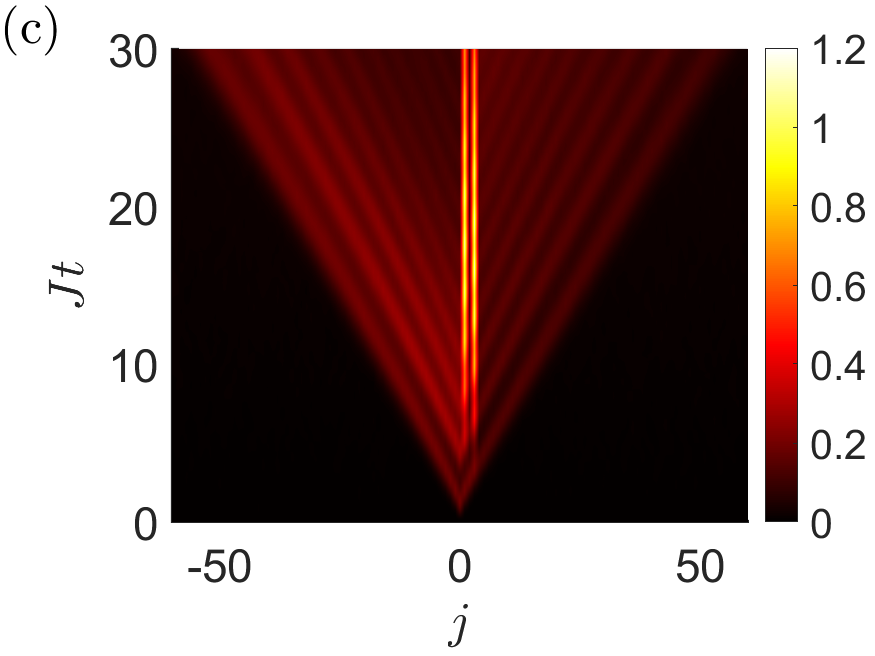}
  \includegraphics[width=0.48\columnwidth]{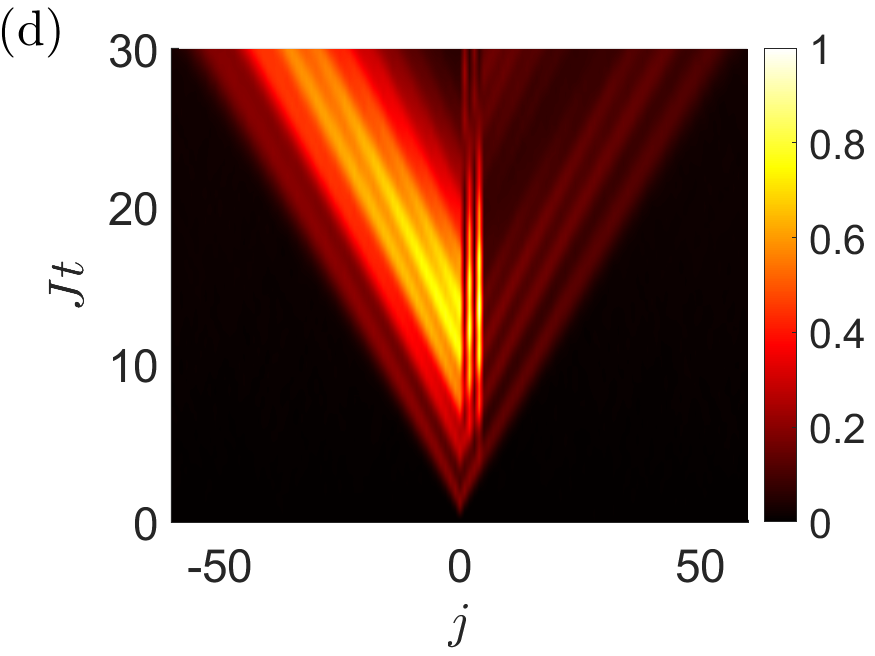}\\
  \caption{(a) The evolution of $\langle S_{T,z} \rangle = \sum_{i=1}^{N_T} \langle \sigma_{T,z}^{(i)} \rangle$ for the fully excited ensemble of TAs with $N_T = 30$. (b) The strength of superradiance $I$ of the TAs as a function of their number of $N_C$, the numerical fitting results are $(\alpha_1,\alpha_2,\alpha_3)=(2.3,0.019,7.4)\times10^{-3}$. (c) and (d) The dynamical evolution of the photons in the waveguide with $N_T = 30$. The parameters are set as $\omega_C=\omega_T=\omega, g = 0.1J, G_1 = 0,\kappa=0.01J$, and $N_C = 10$. For the Dicke superradiance in (a) and (b), we set $G_2 = 0$, while for the other cases, $G_2 = 0.15J$. In these simulations, $\Delta L$ can be arbitrary, and we set $\Delta R = 4$ in (c) and $\Delta R = 5$ in (d), respectively. For all results, we have averaged over $4000$ trajectories.}
  \label{dynamics}
\end{figure}

\emph{Controllable Superradiance}-We first consider the case where the CAs are decoupled from the waveguide, i.e., \(G_1 = G_2 = 0\). For small \(g \ll J\) and \(\omega_T \in (\omega - 2J, \omega + 2J)\), an ensemble of \(N_T\) excited TAs decays with a characteristic rate \(\Gamma_T \propto N_T^2 g^2 /J\), emitting photons into the propagating band of the waveguide, as shown in Figs.~\ref{dynamics}(a) and (b) (blue curves). The \(N_T^2\) factor originates from superradiance due to interference among the atoms.

Next, we consider the effect of the CAs. In the case where CAs couple to the waveguide via one site (small atoms of $G_1=0$), photons emitted by the TAs are reflected by CAs as they propagate along the waveguide, allowing both atomic and photonic evolutions to be controlled by the CAs. The atomic evolution of $S_{T,z}=\sum_i \sigma_{T,z}^{(i)}/2$ in Fig.~\ref{dynamics}(a) shows that the CAs can either accelerate or slow down the emission of the TA's Dicke radiance depending on the distance \(\Delta R = N - n\) between the two ensembles. The superradiance strength, \(I = |d \langle S_{T,z} \rangle / dt|\), at the half-decay time \(T_h\) (where \(\langle S_{T,z} \rangle(T_h) = 0\)), is plotted in Fig.~\ref{dynamics}(b). The scaling \(I \propto N_T^\alpha\) is controlled by the position of the CAs. For \(\Delta R = 4\), \(\alpha \approx 3.2\), and for \(\Delta R = 5\), \(\alpha \approx 1.7\), compared to the standard Dicke superradiance scaling \(\alpha = 2.0\), which can be understood by the atomic correlation. For \(\Delta R = 4\), the TAs experience fractional dissipation or subradiance, preserving some atomic excitation in the long-time limit. The photonic evolution for \(\Delta R = 4\) and \(\Delta R = 5\) is shown in Figs.~\ref{dynamics}(c) and (d), where \(G_1 = 0\) and \(G_2 = 0.15J\). For \(\Delta R = 4\), photons are mostly confined in the sites between TAs and CAs, with a small portion spreading symmetrically along the CRW. For \(\Delta R = 5\), clear chiral superradiance is observed, with more photon intensity radiated to the left side of the atomic regime in the CRW than to the right side.

\begin{figure}
  \includegraphics[width=0.48\columnwidth]{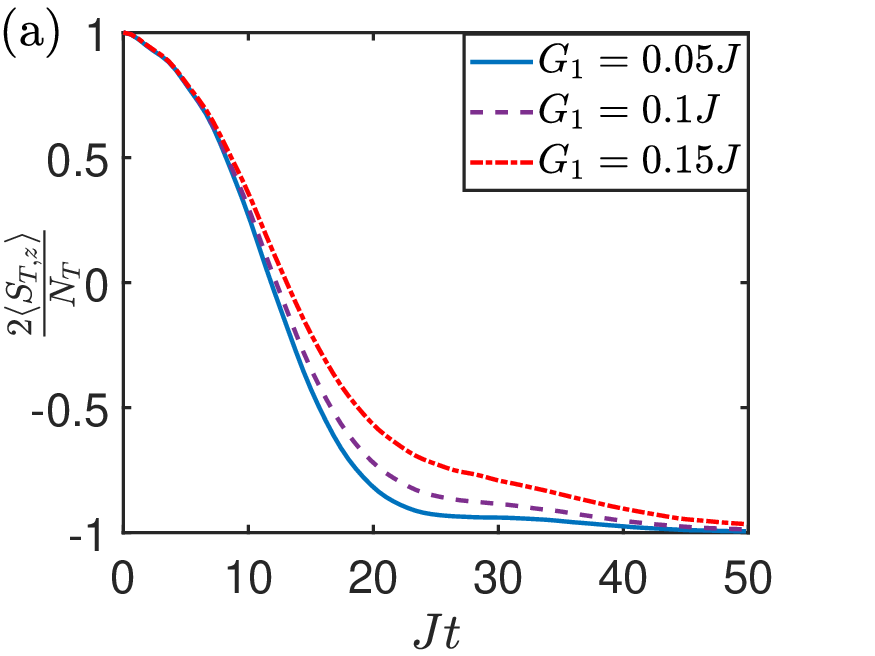}
  \includegraphics[width=0.48\columnwidth]{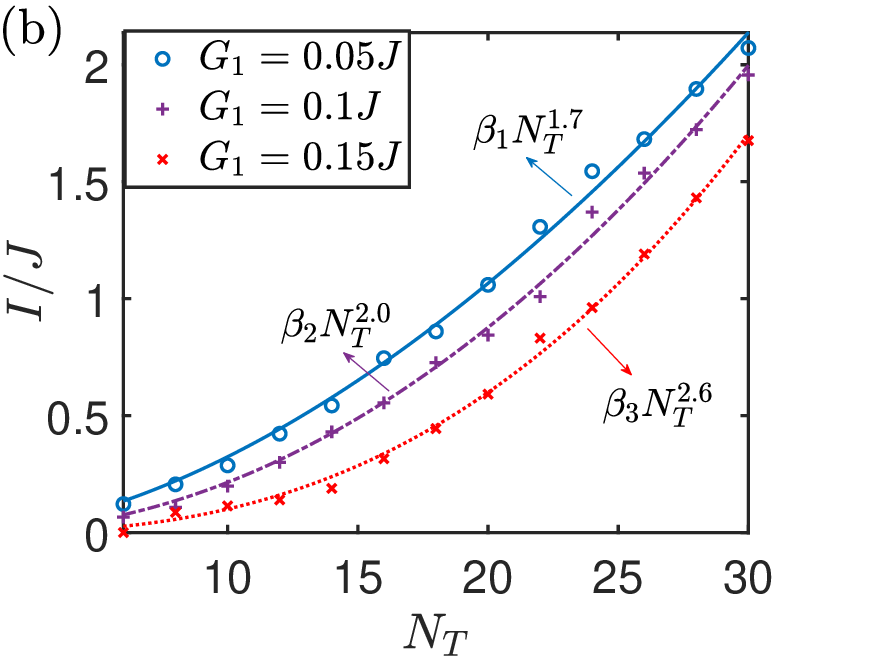}\\
  \includegraphics[width=0.48\columnwidth]{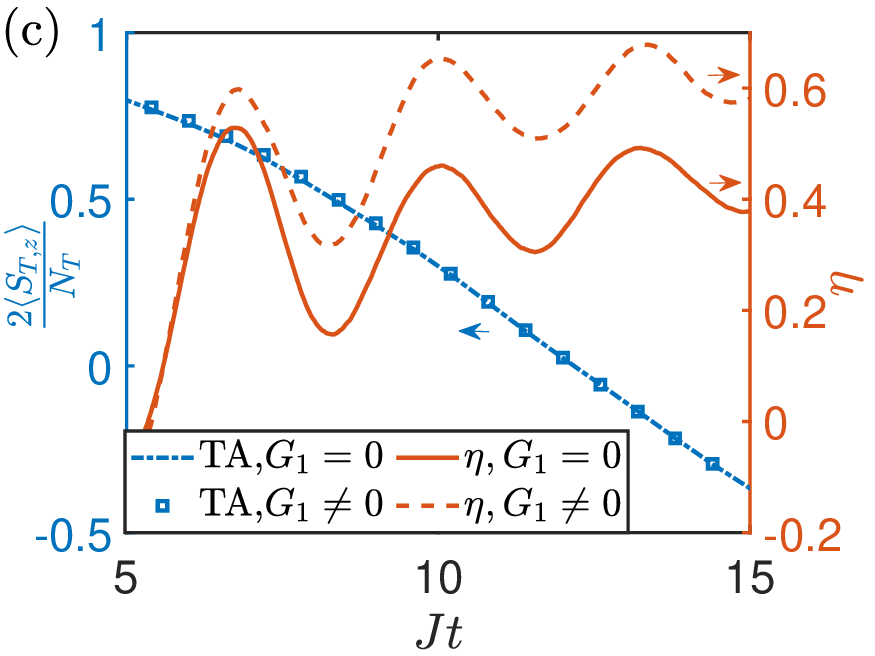}
  \includegraphics[width=0.48\columnwidth]{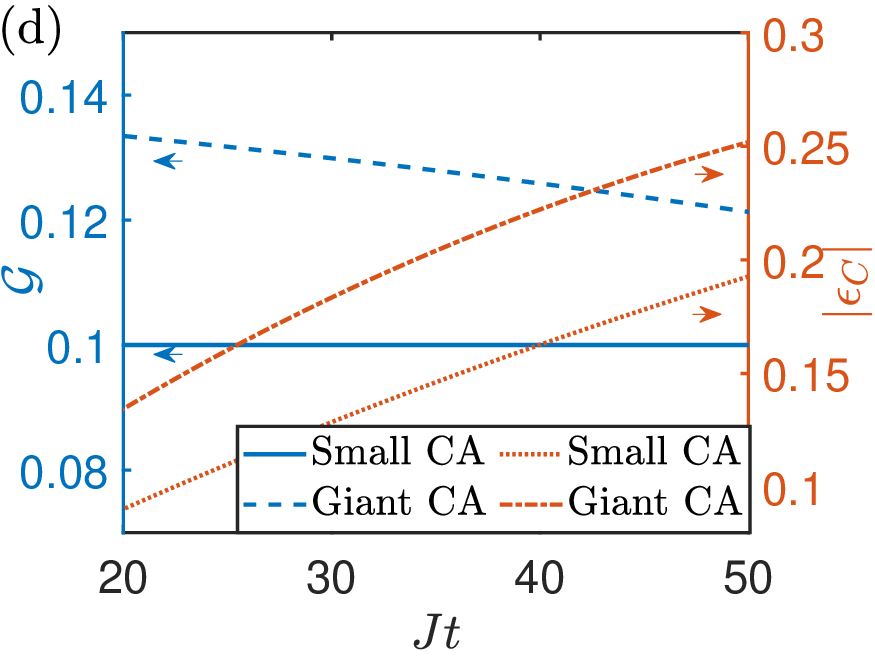}
  \caption{ (a) The evolution of \(\langle S_{T,z}\rangle\) for giant CAs setup. (b) strength of radiance \(I\) for giant CAs setup. (c) The dynamics of TAs and the degree of chirality with small (giant) CAs. (d) \(\mathcal{G}\) and \(|\epsilon_C|\) for small and giant CA setups. The numerical fitting results in (b) are $(\beta_1, \beta_2, \beta_3) = (6.2, 2.0, 0.27) \times 10^{-3}$. In (a), (b) ((c),(d)), we set \(\Delta L=4\), \(\Delta R=5\) (\(\Delta L=2\), \(\Delta R=5\)). For all results in (a), (b) and (c), we have set \(N_T = 30\), \(N_{C}=10\) and averaged over $4000$ trajectories. In (c), for the small (giant) CA setup, we set \(G_1 = 0\ (0.067J)\), \(G_2 = 0.1J\ (0.15J)\) to keep the same dynamics of TAs. In (d), we set \(N_T = 1\), \(N_C = 1\), and for the small (giant) CA setup, we set \(G_1 = 0\ (0.06J)\), \(G_2 = 0.1J\ (0.15J)\) to keep the same dynamics of the TA.  The other parameters are set as \(\omega_T = \omega_C = \omega\), \(g = 0.1J\), \(G_{2}=0.15J\), \(\kappa = 0.01J\), .  }
  \label{giantCA}
\end{figure}

When the small CAs are replaced by the giant ones, the dynamics of the superradiance of the TAs can still be modulated. For example, by considering the configuration with \( \Delta L = 4 \) and \( \Delta R = 5 \), we illustrate the dynamical evolution of \( \langle S_{T,z} \rangle \) in Fig.~\ref{giantCA} (a). This shows that the coupling strength \( G_1 \) between the left leg of the CA and the waveguide serves as a sensitive controller to manipulate the radiance of the TAs. Additionally, similar to the small giant CAs, we investigate the scaling of the strength \( I \) versus \( N_T \) in Fig.~\ref{giantCA} (b). The scaling can be either below or above the Dicke superradiance, depending on the modulation of the coupling strength \( G_1 \). Furthermore, in Fig.~\ref{SM2}(b) of SM~\cite{SM}, we also find that the giant CA can induce chirality in the superradiance.

To quantify the chiral radiance of the TAs, we define the degree of chirality \( \eta \) as
\begin{equation}
\eta = \frac{\langle \hat{a}_{-1}^\dagger \hat{a}_{-1} \rangle - \langle \hat{a}_{N+1}^\dagger \hat{a}_{N+1} \rangle}{\langle \hat{a}_{-1}^\dagger \hat{a}_{-1} \rangle + \langle \hat{a}_{N+1}^\dagger \hat{a}_{N+1} \rangle},
\label{eta}
\end{equation}
which is plotted in Fig.~\ref{giantCA} (c). We observe that the chirality can be achieved with \( \eta \approx 0.4 \) for small CAs \((G_{1}=0)\) and \( \eta \approx 0.6 \) for giant CAs \((G_{1}\neq0)\) at the moment \( Jt = 15 \). This implies that the giant CAs can enhance the degree of chirality compared to the small CA setup. Here, we have appropriately chosen the atom-waveguide coupling strength to ensure the same dynamics (as illustrated by the blue dot-dashed and square curves), and hence the same total photon intensity emitted by the TAs.

\emph{Physical mechanism}-Now, we analytically explain the superradiance of the TAs controlled by the CAs when $N_T\sim N_C$, considering a minimal model with one TA and one CA. Using the Fourier transform \(\hat a_k = \sum_j a_j e^{-ikj}/\sqrt{N_W}\), the Hamiltonian in momentum space and the rotating frame is
\begin{equation}
\hat{H} = \sum_k \omega_k \hat a_k^\dagger \hat a_k + \frac{1}{\sqrt{N_W}} \left( g \hat \sigma_{T,+} e^{-ikn} + g_k \hat \sigma_{C,+} \right) \hat a_k + \text{H.c.}
\end{equation}
where \(\omega_k = -2J \cos k\) is the dispersion relation (with \(\omega_T = \omega_C =\omega\)), and the coupling strength between the CA and the \(k\)-th mode in the waveguide is \(g_k = G_1 + G_2 e^{ikN}\).

Thanks to the conservation of excitations, the single-excitation wave function can be written as
\begin{equation}
\ket{\psi(t)} = \left[ \epsilon_T(t) \hat \sigma_{T,+} + \epsilon_C(t) \hat \sigma_{C,+} + \sum_k c_k(t) \hat a_k^\dagger \right] \ket{g,g, \rm vac}
\label{wavefunction}
\end{equation}
where \(\epsilon_{T(C)}(t)\) is the excitation amplitude for the TA (CA), \(c_k(t)\) is the photonic amplitude for wave vector \(k\), and \(\ket{g,g,\rm vac}\) is the ground state.

Solving the Schr\"{o}dinger equation \(i \partial \ket{\psi(t)}/\partial t = \hat{H} \ket{\psi(t)}\) and eliminating \(c_k\), we obtain the atomic dynamics
\(d\vec{\epsilon}(t)/dt=-M \vec{\epsilon}(t)/(2J)\), where \(\vec{\epsilon}(t) = (\epsilon_T(t), \epsilon_C(t))^T\) and the matrix \(M\) is (see SM~\cite{SM})
\begin{equation}
M = \begin{pmatrix}
g^2 & g \left( G_1 e^{i\phi_L} + G_2 e^{i\phi_R} \right) \\
g \left( G_1 e^{i\phi_L} + G_2 e^{i\phi_R} \right) & G_1^2 + G_2^2 + 2 G_1 G_2 e^{i(\phi_L + \phi_R)}
\end{pmatrix}
\end{equation}
with \(\phi_L = K \Delta L\) and \(\phi_R = K \Delta R\) representing the accumulated phases during photon propagation. Here, \(K = \pi/2\) is the wave vector since both the CA and TA are resonant with the bare resonator in the waveguide. In the SM~\cite{SM}, we have discussed in detail how the eigenvalues of the matrix $M$ correspond to different atomic dynamics for fractional and complete dissipation for \(\Delta R = 4\) and \(\Delta R = 5\), with the BIC being present and absent, respectively, as shown in Fig.~\ref{dynamics}(a).

Furthermore, the chirality of the superradiance can be fundamentally understood through the photonic interference effect. Solving for the photonic amplitude \( c_m(t) = \sum_k c_k e^{imk} / \sqrt{N_W} \) in Eq.~(\ref{wavefunction}), we obtain (see SM~\cite{SM} for details)
\begin{eqnarray}
\begin{split}
c_m(t) = -\frac{i}{2J} \Big[ & g e^{i K R_{mn}} \Theta(t - \tau_{mn}) \epsilon_T(t - \tau_{mn}) \\
& + G_1 e^{i K R_{m0}} \Theta(t - \tau_{m0}) \epsilon_C(t - \tau_{m0}) \\
& + G_2 e^{i K R_{mN}} \Theta(t - \tau_{mN}) \epsilon_C(t - \tau_{mN}) \Big],
\end{split}
\label{CM}
\end{eqnarray}
where \( R_{mn} = |m - n| \) is the distance between coupling sites, \( \tau_{mn} = R_{mn} / (2J) \) is the retardation time, \( \Theta(\cdot) \) is the step function, and the expressions for the complex atomic amplitudes \( (\epsilon_C, \epsilon_T) \) can be found in the SM~\cite{SM}. The degree of chirality \( \eta \) in Eq.~(\ref{eta}) can be extracted from the expressions for \( c_{-1} \) and \( c_{N+1} \). For the small CA setup, we have demonstrated the corresponding phases that reveal the interference effect in Fig.~S4 of the SM~\cite{SM} for \( \Delta R = 4 \) and \( \Delta R = 5 \), without and with chiral superradiance, respectively.

In the giant CA setup (\( G_1 \neq 0 \)), we investigate the fundamental principles underlying chirality enhancement. Neglecting the retardation effect, the degree of chirality can be approximately expressed as (see SM~\cite{SM} for details)
\begin{equation}
\eta \approx \frac{2G_2 \sin \Delta |\epsilon_C|}{g |\epsilon_T|},
\label{etat}
\end{equation}
where \( \Delta = \text{Arg}(\epsilon_T) - \text{Arg}(\epsilon_C) \) represents the phase difference between the TA and CA. In Fig.~\ref{giantCA}(d), we plot the values of \( \mathcal{G} = G_2 \sin \Delta \) and \( |\epsilon_C| \) for both small and giant CA setups. Both values are larger in the giant CA setup, which enhances the chirality of the radiation.

\begin{figure}
  \includegraphics[width=0.48\columnwidth]{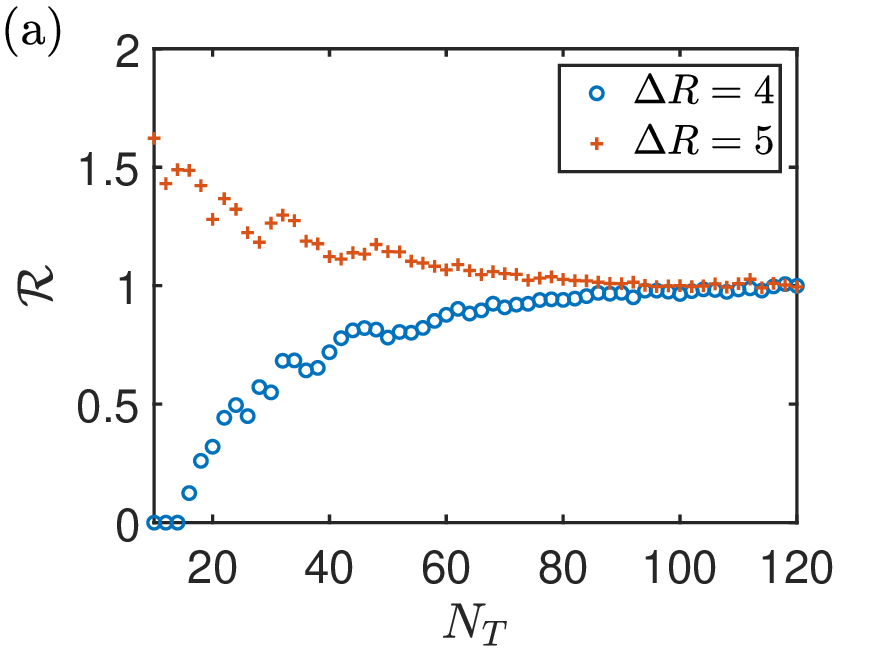}
  \includegraphics[width=0.48\columnwidth]{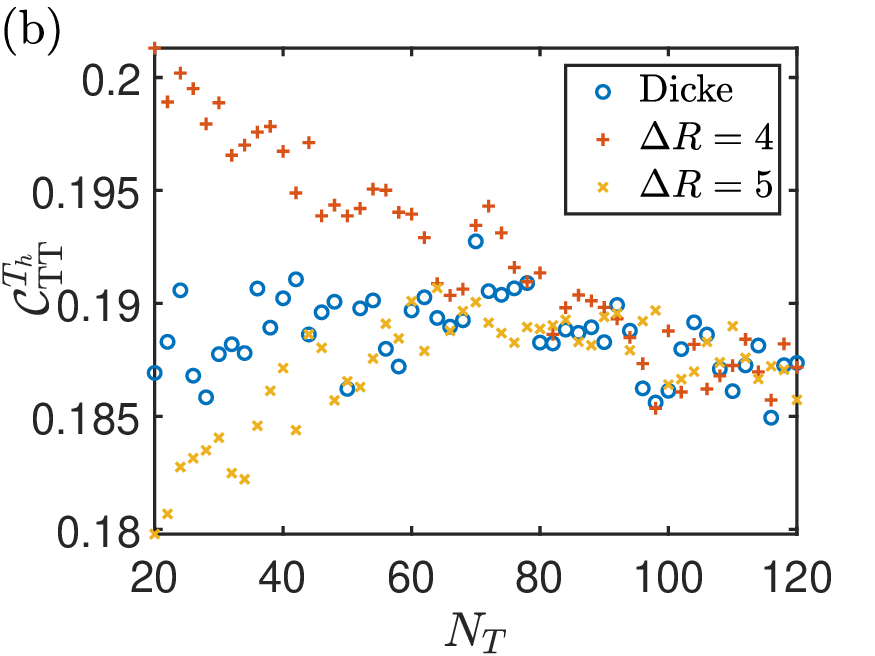}\\
  \includegraphics[width=0.48\columnwidth]{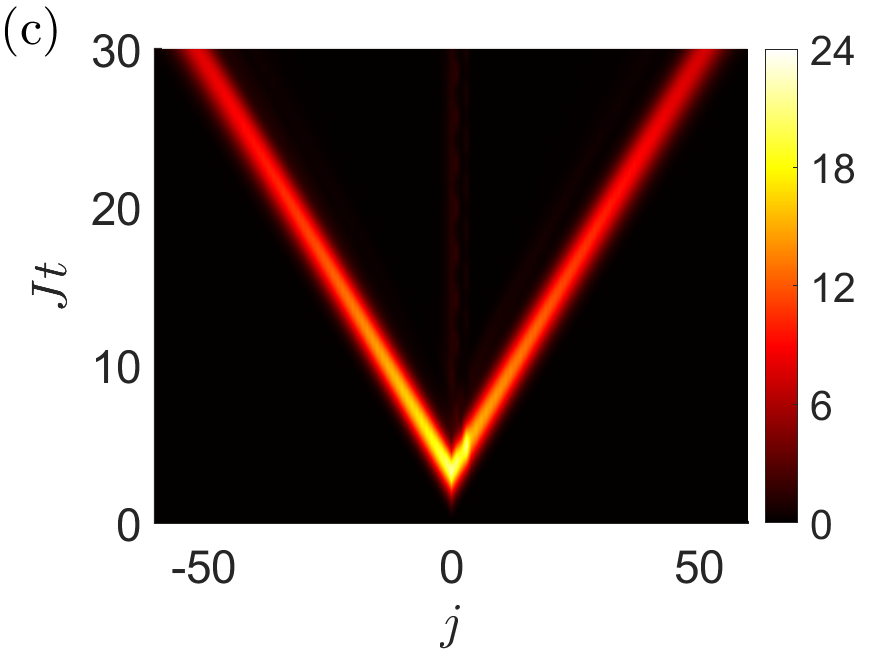}
  \includegraphics[width=0.48\columnwidth]{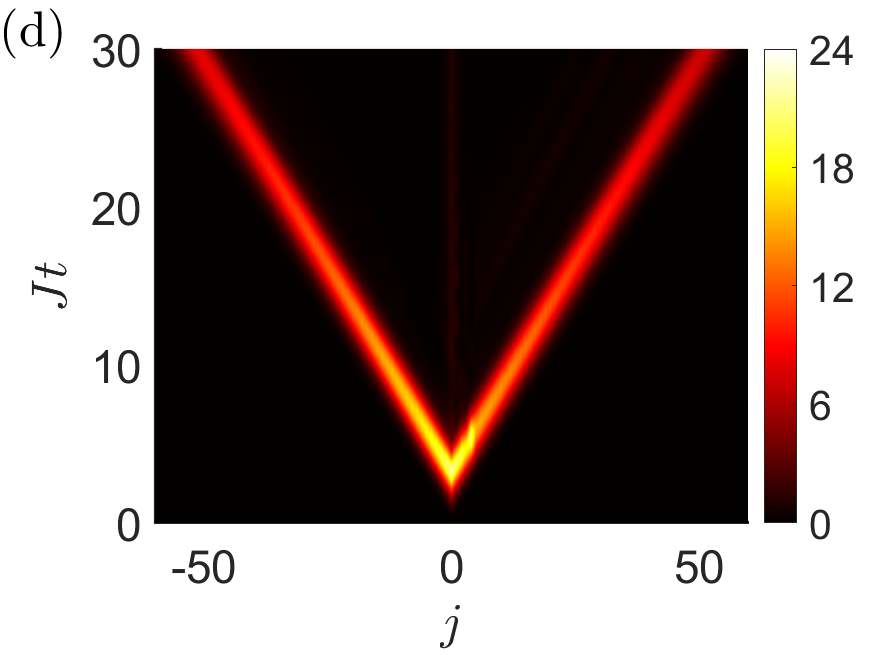}
  \caption{  (a) The ratios \(\mathcal{R}\) for small CAs. (b) The atomic correlation at half-time $T_h$ \((\mathcal{C}_{\rm TT}^{T_h})\) . (c) and (d) The photonic dynamics during the superradiance process with small CAs. The parameters are set as \(\omega_T = \omega_C = \omega\), \(g = 0.1J\), \(G_{1}=0\), \(G_{2}=0.15J\), \(\kappa = 0.01J\), \(N_C = 10\). In (c) and (d), we have set \(N_T=200\), \(\Delta R=4\) and \(\Delta R=5\) respectively. For all results, we have averaged over $4000$ trajectories. }
  \label{largeNT}
\end{figure}

\emph{$N_T\gg N_C$ limit}-We have shown that CAs can engineer the superradiance of TAs when the numbers of CAs and TAs are comparable. The superradiance strength scales either higher or lower  than traditional Dicke superradiance. In Fig.~\ref{largeNT}(a), we plot the strength ratio \(\mathcal{R} = I / I_{\rm Dicke}\) as a function of \(N_T\). For \(N_T \lesssim 5N_C\), the dependence of \(\mathcal{R}\) on \(N_T\) and \(\Delta R\) shows that the superradiance of the TAs can be effectively controlled by the CAs. This agrees with the scaling modulation as shown in Fig.~\ref{dynamics}(b). However, as \(N_T\) increases further, the ratio \(\mathcal{R}\) gradually approaches $1$ for \(N_T \gg N_C\), indicating a recovery of standard Dicke superradiance, which obeys the \(N_{T}^2\) scaling.

This behavior can be explained by the average two-atom correlation in TAs at half-time \(T_h\), \(\mathcal{C}_{\rm TT}^{T_h} = \sum_{i \neq j} \langle \sigma_{T,+}^{(i)} \sigma_{T,-}^{(j)} \rangle / (N_T^2 - N_T)\), which is shown in Fig.~\ref{largeNT}(b). For small \(N_T\), the correlations depend on \(\Delta R\) and \(N_{T}\). The correlation for \(\Delta R=5\) approaches the standard Dicke superradiance compared to \(\Delta R=4\), coincide with the difference of scaling with Dicke superradiance given by Fig.~\ref{dynamics}(b).  However, when the number of TAs becomes large enough (\(N_T > 10N_C\)), the ratio \(\mathcal{R} \approx 1\), and the correlations become independent of the parameters, resulting in identical superradiance strength for different values of \(\Delta R\).

Additionally, for \(N_T = 200, N_C = 10\), we investigate the photonic dynamics in Fig.~\ref{largeNT}(c,d) under the same parameters as in Fig.~\ref{dynamics}(c,d). The results show that the photons are emitted symmetrically in both directions, without photon bounding or chiral radiation, further indicating the recovery of standard Dicke superradiance.

\begin{figure}
  \includegraphics[width=0.48\columnwidth]{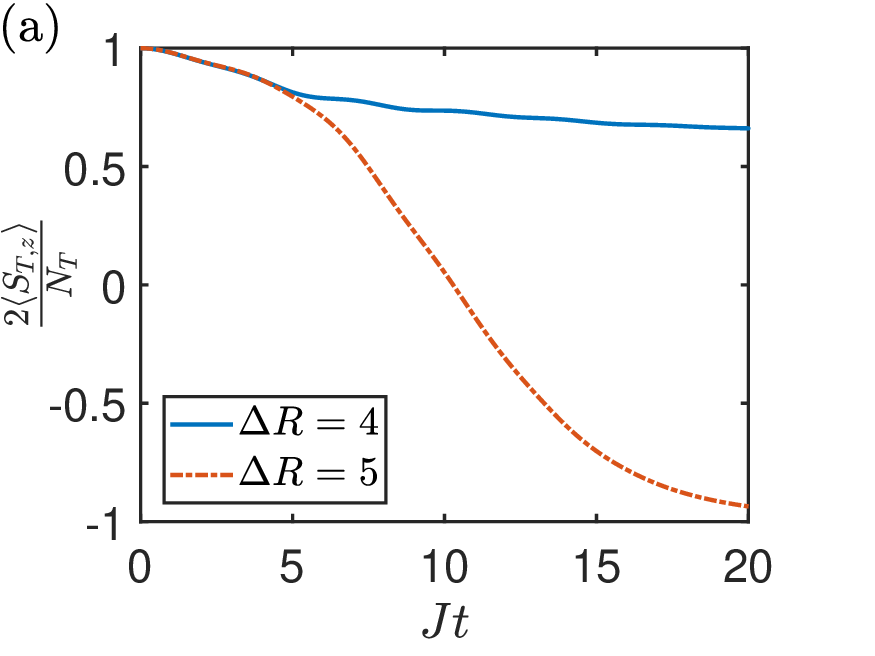}
  \includegraphics[width=0.48\columnwidth]{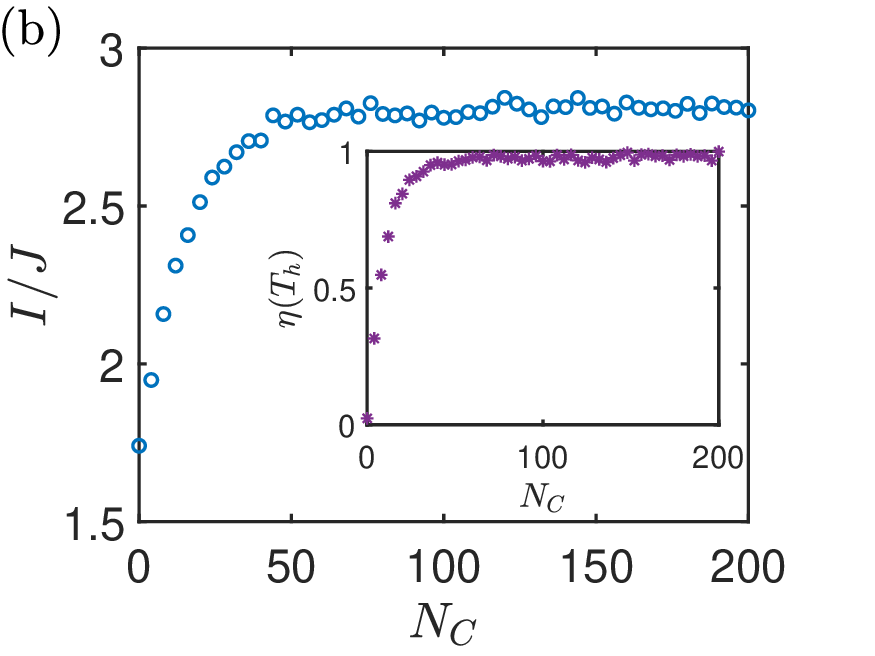}\\
  \includegraphics[width=0.48\columnwidth]{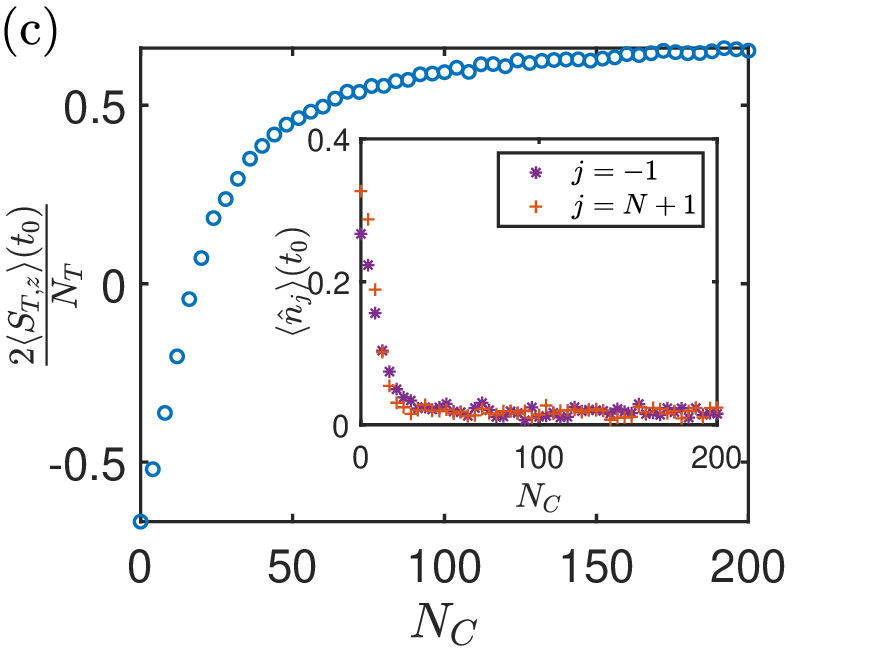}
  \includegraphics[width=0.48\columnwidth]{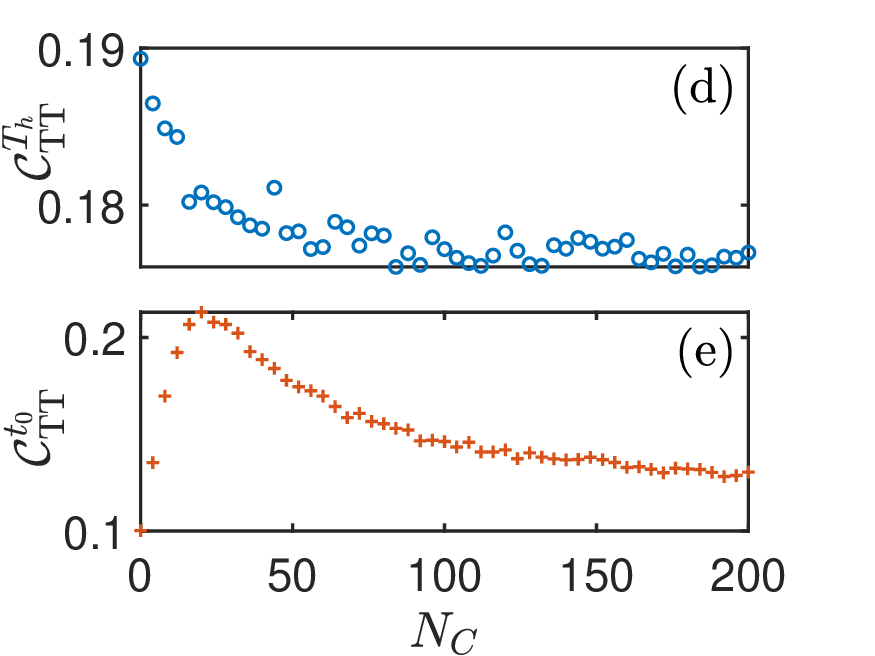}
  \caption{  (a) and (c) The dynamics of TA in the small CAs setup. (b) The superradiance strength \(I\), and the degree of chirality \(\eta\) (inset). (c) The dynamics of CAs and the photonic dynamics in the $-1$th and $N+1$th resonators in the waveguide.  (d) The atomic correlation \(\mathcal{C}_{TT}^{T_h}\) at half-time \(T_h\) for $\Delta R=5$. (e) The atomic correlation \(\mathcal{C}_{TT}^{t_0}\) at $Jt_0=20$ for $\Delta R=4$. The parameters are set as \(\omega_T = \omega_C = \omega\), \(g = 0.1J\), \(G_{1}=0\), \(G_{2}=0.15J\), \(\kappa = 0.01J\), and \(N_T = 30\). For all results, we have averaged over $4000$ trajectories. }
  \label{largeNC}
\end{figure}

\emph{$N_T\ll N_C$ limit}-In the opposite limit of \(N_T \ll N_C\), the atomic dynamics in Fig.~\ref{largeNC}(a) show complete (fractional) dissipation when the BIC is absent (present) for \(\Delta R = 5\,(4)\) in the small CA setup. For \(\Delta R = 5\), we further investigate the radiance strength \(I = |d\langle S_{T,z}\rangle / dt|\) at half-time \(T_h\) (approximately obeys \(I\propto N_T^2\) for \(N_{T}\ll N_{C}\)) as a function of \(N_C\) in Fig.~\ref{largeNC}(b), which shows a saturation effect with increasing \(N_C\). Along with the strength saturation, the degree of chirality \(\eta \to 1\) in the inset subfigure, indicating nearly perfect chiral superradiance. In the case of \(\Delta R = 4\), as shown in Fig.~\ref{largeNC}(c), where the evolution time \(t_0\) satisfies \(Jt_0 = 20\), most of the TAs are trapped in the excited state with \(2\langle S_z \rangle / N_T \approx g / G_2\). This shows that large \(N_C\) further enhances the ability of the BIC to protect the TAs from complete dissipation. Meanwhile, the photon distribution at the \(-1\)th and \((N+1)\)th sites is nearly zero for large \(N_C\) (see inset subfigure), implying that the photon is trapped inside the atomic regime, consistent with the BIC physics.

The above \(N_C\)-independent (for large $N_C$) behaviors can also be understood through the correlations. As shown in Figs.~\ref{largeNC}(d) and (e), we plot \(\mathcal{C}_{\rm TT}^{T_h}\) for \(\Delta R = 5\) and \(\mathcal{C}_{\rm TT}^{t_0}\) for \(\Delta R = 4\), respectively. For the case of \(N_C \gg N_T\), we find that both atomic correlations reach steady values that are independent of \(N_C\). This explains why we observe the saturation effect in Figs.~\ref{largeNC}(b) and (c).

\emph{Physical realizations}-Our model could be explored in the superconducting circuit QED system. A CRW consisting of a $42$-unit-cell array of microwave resonators has been realized~\cite{XZ2023}, with photonic hopping rates on the order of $100$ MHz. The coupling between a single transmon qubit and the resonator is of the same magnitude~\cite{EK2021}. Furthermore, giant atoms have been implemented using transmons~\cite{BK2020,AM2021} or magnons~\cite{ZQ2022}, and the superradiance of giant atom ensembles has been investigated in recent studies~\cite{AL2024}. In our work, the ensemble of TAs can be realized using Rydberg atoms trapped above the resonator array~\cite{DP2008,SD2012}. Alternatively, the CRW could be replaced with a transmission line or optical fiber supporting a linear dispersion relation~\cite{JD2013s,SK2017}, simplifying experimental implementation.

\emph{Conclusions}-We have demonstrated the superradiance of an ensemble of two-level atoms in a 1D photonic waveguide. By engineering another distant atomic ensemble that couples to the same waveguide, we can control the superradiance behavior on demand, with the superradiance scaling being either lower or higher than that of standard Dicke superradiance. Additionally, we have realized chiral superradiance induced by interference effects, where the photon emission strength in one direction is significantly stronger than in the opposite direction. We also find that using giant atoms as  control atoms enhances the degree of chirality compared to using smaller atoms, even when the superradiance strength is the same.

Beyond the traditional mean-field approximation and master equation, we apply the DTWA technique to capture both atomic and photonic dynamics, especially when the number of participating atoms is large but far from the thermodynamic limit. Our work thus opens new avenues for exploring the controllability of superradiance across various many-body platforms.

\emph{Acknowledgments}-We thank Prof. Xin-You L\"{u} for the helpful discussion. This work is supported by the Science and Technology Development Project of Jilin Province (Grant No. 20230101357JC), National Science Foundation of China (Grant No. 12375010) and the Innovation Program for Quantum Science and Technology (No. 2023ZD0300700)

    %%%%%%%%%% Merge with supplemental materials %%%%%%%%
\clearpage
\newpage
\onecolumngrid

%%%%%%%%%% Prefix a "S" to all equations, figures, tables and reset the counter %%%%%%%%%%
\newcommand\specialsectioning{\setcounter{secnumdepth}{-2}}
\setcounter{equation}{0} \setcounter{figure}{0}

\setcounter{table}{0}
\renewcommand{\theequation}{S\arabic{equation}}
\renewcommand{\thefigure}{S\arabic{figure}}
\renewcommand{\bibnumfmt}[1]{[S#1]}
\renewcommand{\citenumfont}[1]{S#1}
\renewcommand\thesection{S\arabic{section}}
%%%%%%%%%% Prefix a "S" to all equations, figures, tables and reset the counter %%%%%%%%%%
\renewcommand{\baselinestretch}{1.2}

%\renewcommand{\theequation}{S\arabic{equation}}

%%%%%%%%%%%%%%%%%%%%%%%%%%%%%%%%%%%%%%%%%%%%%%%%%%%%%%%%%%%%%%%%%

\setcounter{page}{1}\setcounter{secnumdepth}{3} \makeatletter
\begin{center}
	{\Large \textbf{ Supplemental Material for\\
			``Controllable superradiance scaling in photonic waveguide"}}
\end{center}

\begin{center}
	Xiang Guo and Zhihai Wang$^{*}$
\end{center}
\begin{minipage}[]{16cm}
	\small{\it
		\centering Center for Quantum Sciences and School of Physics, Northeast Normal University, Changchun 130024, China}
\end{minipage}

\bigskip

This supplementary material is divided into two sections. In Sec.~\ref{s1}, we provide the details of the discrete truncated Wigner approximation (DTWA), which captures both atomic and photonic dynamics during the superradiance process of the target atoms (TAs), manipulated by the control atoms (CAs). In Sec.~\ref{s2}, we derive the dynamics of the minimal model consisting of one TA and one CA, which explains the chiral and bound state in the continuum (BIC) physics.

\section{Discrete Truncated Wigner Approximation}\label{s1}

In the main text, we have investigated the superradiance of the target atoms subject to a structured reservoir composed of a coupled resonator waveguide, considering both atomic and photonic dynamical evolution. When the number of TAs is \(N_T\), the number of CAs is \(N_C\), and \(N_W\) resonators are considered, the dimension of the Hilbert space for the whole system is \(d = (N_C + 1) \times (N_T + 1) \times N_0^{N_W}\) (with \(N_0\) being the photon number cutoff in each resonator), even when accounting for the atomic exchange symmetry. Therefore, the numerical solution of the master equation (Eq.~(\ref{master}) in the main text) becomes impractical for large \(N_C\), \(N_T\), \(N_W\), and \(N_0\), even though they are far from the thermodynamic limit.

To tackle the above issue, we adopt DTWA~\cite{JS2015s}, which transforms the master equation into Fokker-Planck-type equations of motion for the atomic and photonic amplitudes in the waveguide. To this end, we begin with the Heisenberg equation based on the Hamiltonian in Eq.~(\ref{Hamil}), that is

\begin{eqnarray}
\frac{d\hat\sigma_{T,x}^{(i)}}{dt}&=&-\omega_{T}\hat\sigma_{T,y}^{(i)}+ig(\hat\sigma_{T,z}^{(i)}\hat a_{n}-\hat a_{n}^{\dagger}\hat\sigma_{T,z}^{(i)}),\,(i=1,2\dots,N_{T}),\\
\frac{d\hat\sigma_{T,y}^{(i)}}{dt}&=&\omega_{T}\hat\sigma_{T,x}^{(i)}-g(\hat\sigma_{T,z}^{(i)}\hat a_{n}+\hat a_{n}^{\dagger}\hat\sigma_{T,z}^{(i)}),\\
\frac{d\hat\sigma_{T,z}^{(i)}}{dt}&=&g(\hat\sigma_{T,y}^{(i)}\hat a_{n}+\hat a_{n}^{\dagger}\hat\sigma_{T,y}^{(i)})-ig(\hat\sigma_{T,x}^{(i)}\hat a_{n}-\hat a_{n}^{\dagger}\hat\sigma_{T,x}^{(i)}),\\
\frac{d\hat\sigma_{C,x}^{(j)}}{dt}&=&-\omega_{C}\hat\sigma_{C,y}^{(j)}+iG_{1}(\hat\sigma_{C,z}^{(j)}\hat a_{0}-\hat a_{0}^{\dagger}\hat\sigma_{C,z}^{(j)})+iG_{2}(\hat\sigma_{C,z}^{(j)}\hat a_{N}-\hat a_{N}^{\dagger}\hat\sigma_{C,z}^{(j)})\,(j=1,2\dots,N_{C}),\\
\frac{d\hat\sigma_{C,y}^{(j)}}{dt}&=&\omega_{C}\hat\sigma_{C,x}^{(j)}-G_{1}(\hat a_{0}^{\dagger}\hat\sigma_{C,z}^{(j)}+\hat\sigma_{C,z}^{(j)}\hat a_{0})-G_{2}(\hat a_{N}^{\dagger}\hat\sigma_{C,z}^{(j)}+\hat\sigma_{C,z}^{(j)}\hat a_{N}),\\
\frac{d\hat\sigma_{C,z}^{(j)}}{dt}&=&G_{1}[\hat a_{0}^{\dagger}(\hat\sigma_{C,y}^{(j)}+i\hat\sigma_{C,x}^{(j)})+(\hat\sigma_{C,y}^{(j)}-i\hat\sigma_{C,x}^{(j)})\hat a_{0}]+G_{2}[\hat a_{N}^{\dagger}(\hat\sigma_{C,y}^{(j)}+i\hat\sigma_{C,x}^{(j)})+(\hat\sigma_{C,y}^{(j)}-i\hat\sigma_{C,x}^{(j)})\hat a_{N}],\\
\frac{d\hat a_{m}}{dt}&=&-i\omega_{c}\hat a_{m}+iJ(\hat a_{m-1}+\hat a_{m+1}),\,(m\neq n,0,N),\\
\frac{d\hat a_{0}}{dt}&=&-i\omega_{c}\hat a_{0}+iJ(\hat a_{-1}+\hat a_{1})-\frac{iG_{1}}{2}\sum_{j=1}^{N_{C}}
(\hat\sigma_{C,x}^{(j)}-i\hat\sigma_{C,y}^{(j)}),\\
\frac{d\hat a_{N}}{dt}&=&-i\omega_{c}\hat a_{N}+iJ(\hat a_{N+1}+\hat a_{N-1})-\frac{iG_{2}}{2}\sum_{j=1}^{N_{C}}
(\hat\sigma_{C,x}^{(j)}-i\hat\sigma_{C,y}^{(j)}),\\
\frac{d\hat a_{n}}{dt}&=&-i\omega_{c}\hat a_{n}+iJ(\hat a_{n+1}+\hat a_{n-1})-\frac{ig}{2}\sum_{i=1}^{N_{T}}(\hat\sigma_{T,x}^{(i)}-
i\hat\sigma_{T,y}^{(i)}).
\end{eqnarray}

Next, we replace the operators by their mean values, reducing the dimension to \(\tilde{d} = 3(N_C + N_T) + N_W\), which scales linearly with \(N_C\), \(N_T\), and \(N_W\), making the calculation computationally inexpensive. However, in the averaging process, correlations and fluctuations are excluded. Fortunately, in the DTWA treatment, quantum fluctuations are considered to the lowest order by using Monte Carlo samples and by coupling the resonators to white noise processes, which generate the quantum correlations.

Subsequently, the above equations can be converted to
\begin{eqnarray}
dT_{x}^{(i)}&=&-\omega_{T}T_{y}^{(i)}dt
-2g\text{Im}[\alpha_{n}]T_{z}^{(i)}dt,\,(i=1,2\dots,N_{T}),\label{TX}\\
dT_{y}^{(i)}&=&\omega_{T}T_{x}^{(i)}dt-2g\text{Re}[\alpha_{n}]T_{z}^{(i)}dt,\\
dT_{z}^{(i)}&=&2g\text{Re}[\alpha_{n}]T_{y}^{(i)}dt+2g\text{Im}[\alpha_{n}]
T_{x}^{(i)}dt,\\
dC_{x}^{(j)}&=&-\omega_{C}C_{y}^{(j)}dt+2G_{1}\text{Im}
[\alpha_{0}^{*}]C_{z}^{(j)}dt+2G_{2}\text{Im}[\alpha_{N}^{*}]C_{z}^{(j)}dt,
(j=1,2\dots,N_{C}),\\
dC_{y}^{(j)}&=&\omega_{C}C_{x}^{(j)}dt-2G_{1}
\text{Re}[\alpha_{0}^{*}]C_{z}^{(j)}dt-2G_{2}\text{Re}[\alpha_{N}^{*}]C_{z}^{(j)}dt,\\
dC_{z}^{(j)}&=&2G_{1}(\text{Re}[\alpha_{0}^{*}]C_{y}^{(j)}
-\text{Im}[\alpha_{0}^{*}]C_{x}^{(j)})dt+2G_{2}(\text{Re}[\alpha_{N}^{*}]
C_{y}^{(j)}-\text{Im}[\alpha_{N}^{*}]C_{x}^{(j)})dt,\\
d\alpha_{m}&=&-i\omega_{c}\alpha_{m}dt+iJ(\alpha_{m-1}+\alpha_{m+1})dt
-\kappa\alpha_{m}dt+\sqrt{\frac{\kappa}{2}}(dw_{1,m}+idw_{2,m}),\,(m\neq n,0,N),\\
d\alpha_{0}&=&-i\omega_{c}\alpha_{0}dt+iJ(\alpha_{1}+\alpha_{-1})dt
-\kappa\alpha_{0}dt-\frac{iG_{1}}{2}\sum_{j=1}^{N_{C}}(C_{x}^{(j)}
-iC_{y}^{(j)})dt+\sqrt{\frac{\kappa}{2}}(dw_{1,0}+idw_{2,0}),\\
d\alpha_{N}&=&-i\omega_{c}\alpha_{N}dt+iJ(\alpha_{N-1}+\alpha_{N+1})dt
-\kappa\alpha_{N}dt-\frac{iG_{2}}{2}\sum_{j=1}^{N_{C}}(C_{x}^{(j)}-iC_{y}^{(j)})
dt+\sqrt{\frac{\kappa}{2}}(dw_{1,N}+idw_{2,N}),\\
d\alpha_{n}&=&-i\omega_{c}\alpha_{n}dt+iJ(\alpha_{n-1}+\alpha_{n+1})dt-\kappa\alpha_{n}dt-\frac{ig}{2}\sum_{i=1}^{N_{T}
}(T_{x}^{(i)}-iT_{y}^{(i)})dt+\sqrt{\frac{\kappa}{2}}(dw_{1,n}+idw_{2,n}),
\label{AN}
\end{eqnarray}
where \(T_{\gamma}^{(i)} = \langle \sigma_{T,\gamma}^{(i)} \rangle\), \(C_{\gamma}^{(i)} = \langle \sigma_{C,\gamma}^{(i)} \rangle\)\,(\(\gamma = x, y, z\)), and \(\alpha_{i} = \langle a_i \rangle\). Here, \(dw_{1,l}\) and \(dw_{2,l}\) are real-valued classical noise terms for the \(l\)th resonator, which satisfy \(\langle dw_{i,n}dw_{j,m} \rangle = \delta_{i,j}\delta_{n,m}dt\). In the realistic simulation~\cite{JS2015s,JH2022s}, the initial conditions are randomly drawn from one of eight configurations
\begin{equation}
(T_x^{(i)}, T_y^{(i)}, T_z^{(i)}) = (\pm1, \pm1, 1), \,
(C_x^{(i)}, C_y^{(i)}, C_z^{(i)}) = (\pm1, \pm1, -1),
\end{equation}
each occurring with equal probability. This represents that all the TAs (CAs) are initially in the excited (ground) states. For the photonic counterpart, the initial values are taken from a Gaussian distribution centered at zero. Using these initial conditions, the dynamical equations of (\ref{TX})-(\ref{AN}) can be solved.

Repeating the above processes for \(N_t \gg 1\) times, the expected values (symmetrically ordered) can be calculated by averaging all trajectories as
\begin{equation}
\langle \hat{\sigma}_{A,\gamma}^{(i)} \rangle \simeq \frac{1}{N_t} \sum_{m=1}^{N_t} A_{\gamma,m}^{(i)}, \,
\langle \{\hat{\sigma}_{A,\gamma}^{(i)} \hat{\sigma}_{B,\beta}^{(j)}\}_{\rm sym} \rangle \simeq \frac{1}{N_t} \sum_{m=1}^{N_t} \left(A_{\gamma,m}^{(i)} B_{\beta,m}^{(j)}\right), \,
\langle \{(\hat{a}_{n}^\dagger)^k (\hat{a}_{m})^l\}_{\rm sym} \rangle \simeq \frac{1}{N_t} \sum_{p=1}^{N_t} \left(\alpha^*_{n,p}\right)^k \alpha_{n,p}^l.
\end{equation}
Here, \(\{\cdot\}_{\rm sym}\) denotes the symmetrically ordered operator product, and the subscripts \(m\) (\(p\)) on the right-hand side represent the values along the \(m\)th (\(p\)th) trajectory.

\begin{figure}
  \includegraphics[width=0.48\columnwidth]{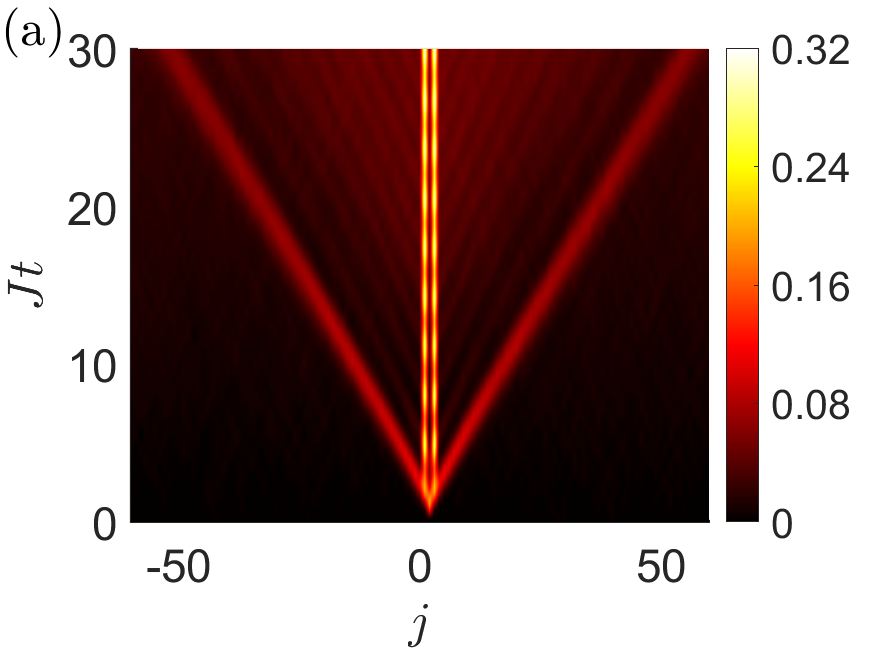}
  \includegraphics[width=0.48\columnwidth]{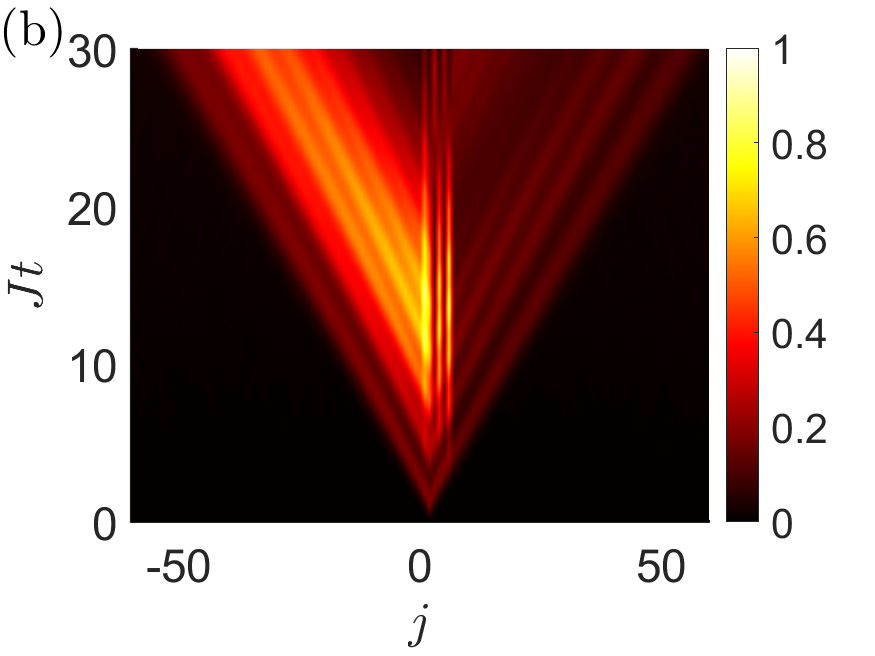}\\
  \caption{The dynamics of photon for giant CAs setup. The parameters are set as \(\omega_T = \omega_C = \omega\), \(g = 0.1J\), \(G_{2}=0.15J\), \(\kappa = 0.01J\), \(N_T = 30\), \(N_{C}=10\). (a) \(G_{1}=0.15J\), \(\Delta L=2\), \(\Delta R=2\) and (b) \(G_{1}=0.067J\), \(\Delta L=2\), \(\Delta R=5\). For all results, we have averaged over $4000$ trajectories. }
  \label{SM2}
\end{figure}

By use of the DTWA, we have demonstrated the atomic and photonic dynamics in Fig.~\ref{dynamics} considering that the CAs are served by the small atoms. As for the giant CAs, we plot the atomic population and strength of superradiance in Figs.~\ref{giantCA}(a) and (b). We observe that, for the fixed $\Delta R$ and $\Delta L$, the coupling strength between the left leg of the CA also plays as a sensitive controller to manipulate the scaling the TAs' superradiance.  Furthermore, for the photonic counterpart, the results in Figs.~\ref{SM2} (a) and (b) show that the photon can be either trapped inside the atom's regime or propagated with chirality.

\section{Physical mechanism in minimal model of one TA and one CA}
\label{s2}

To understand the underlying physical mechanism behind the dynamics during the superradiance of TAs, we adopt a minimal model composed of one TA and one CA. In momentum space, the Hamiltonian of the minimal model is expressed as
\begin{equation}
\hat{H} = \sum_{k} \omega_k \hat{a}_{k}^{\dagger} \hat{a}_{k} + \frac{g}{\sqrt{N_c}} \sum_{k} \left( \hat{\sigma}_{T,+} \hat{a}_{k} e^{-ikn} + \hat{a}_{k}^{\dagger} \hat{\sigma}_{T,-} e^{ikn} \right) + \frac{1}{\sqrt{N_c}} \sum_{k} \left( g_{k} \hat{\sigma}_{C,+} \hat{a}_{k} + g_{k}^{*} \hat{a}_{k}^{\dagger} \hat{\sigma}_{C,-} \right),
\end{equation}
where \(\omega_k = -2J \cos k\) is the dispersion relation of the waveguide, and \(g_{k} = G_{1} + G_{2} e^{ikN}\) represents the coupling strength between the CA and the waveguide. The atomic and resonator frequencies are set to be equal, consistent with the main text.

\begin{figure}
  \includegraphics[width=0.48\columnwidth]{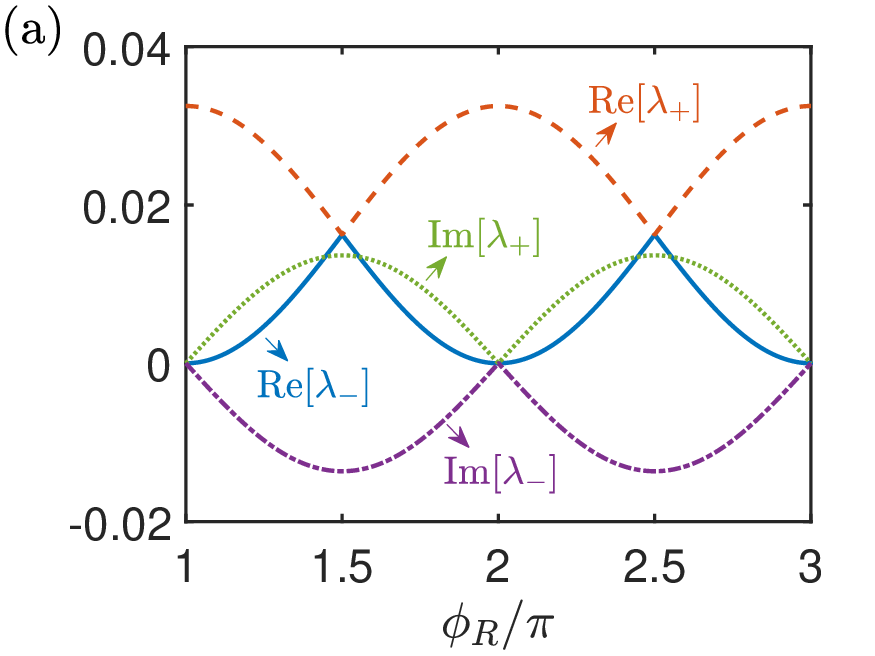}
  \includegraphics[width=0.48\columnwidth]{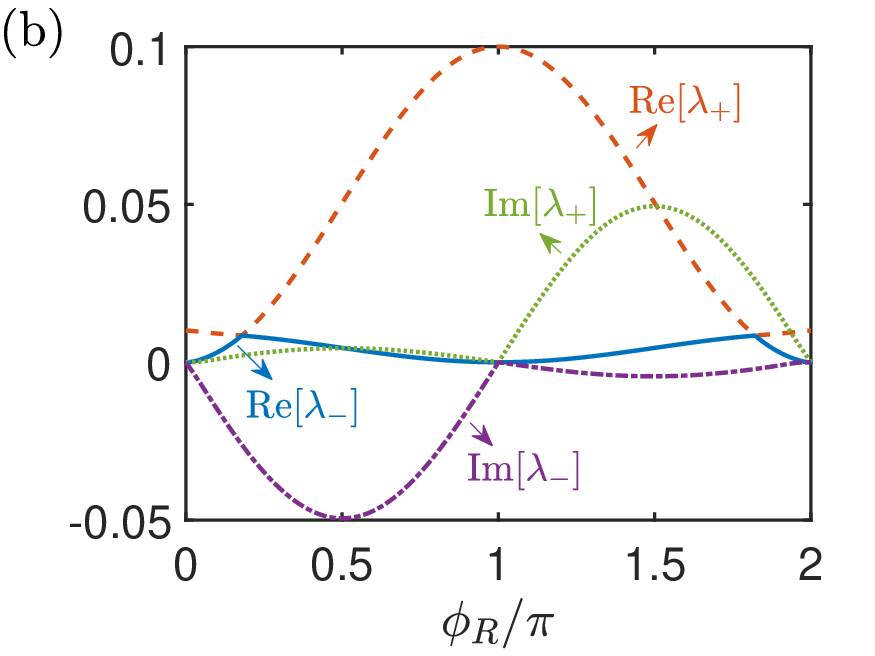}\\
  \includegraphics[width=0.48\columnwidth]{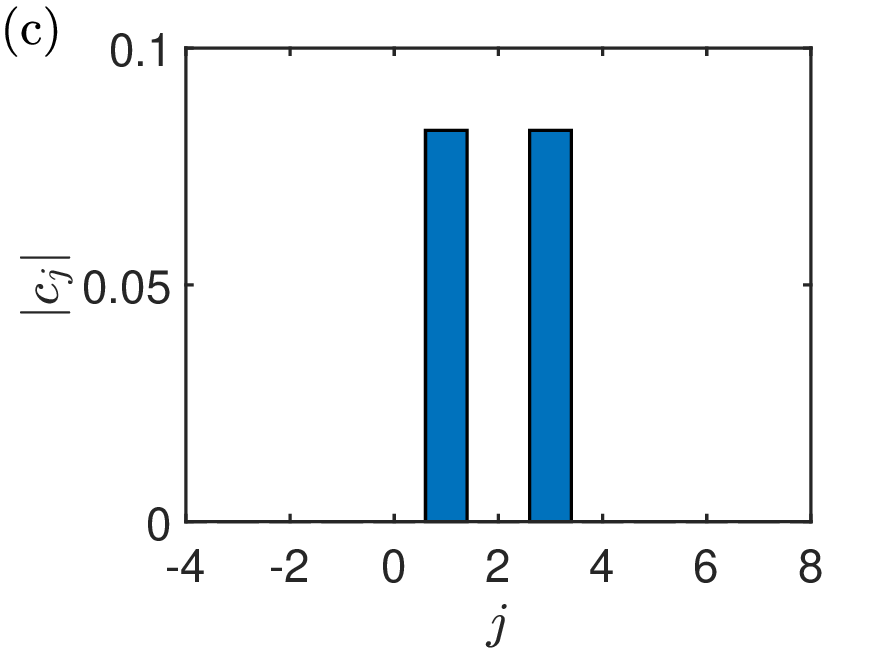}
  \includegraphics[width=0.48\columnwidth]{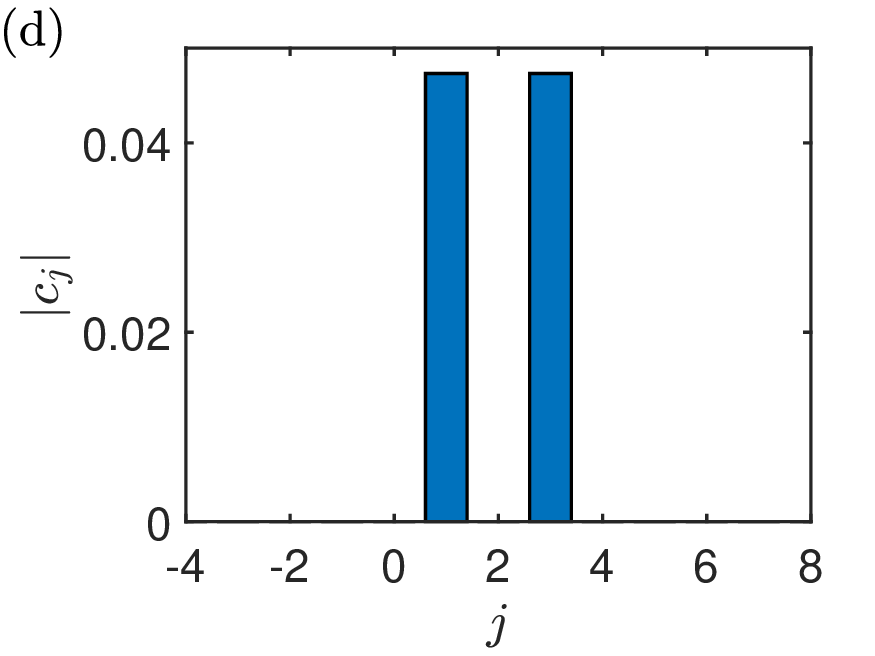}
  \caption{ (a) and (b) The eigenvalues \(\lambda_{\pm}\) of the matrix \(M\) for small and giant CA, respectively. (c) and (d) The photonic distribution of the BIC for small and giant CA, respectively. The parameters are set as \(\omega_T = \omega_C = \omega\), \(g = 0.1J\), \(G_{2}=0.15J\), \(N_T = 1\), \(N_{C}=1\). (a) \(G_{1}=0\), (b)\(G_{1}=0.15J\),\(\phi_L=\pi\), (c) \(G_{1}=0\), \(\phi_R=2\pi\) and (d) \(G_{1}=0.15J\), \(\phi_L=\pi\), \(\phi_R=\pi\).}
  \label{SM3}
\end{figure}

In the single-excitation subspace, the wavefunction for the atom-waveguide coupling system is assumed to be
\begin{equation}
\ket{\psi(t)} = \left( \epsilon_{T}(t) \hat{\sigma}_{T,+} + \epsilon_{C}(t) \hat{\sigma}_{C,+} + \sum_{k} c_{k}(t) \hat{a}_{k}^{\dagger} \right) \big| g, g, \mathrm{vac} \big\rangle,
\end{equation}
where \(\epsilon_{T}\) and \(\epsilon_{C}\) are the excitation amplitudes of the TA and CA, respectively, and \(c_{k}\) denotes the probability amplitude of a photon with wavevector \(k\) in the waveguide. Solving the Sch\"{o}dinger equation \(i \partial \ket{\psi(t)}/\partial t = \hat{H} \ket{\psi(t)}\), we obtain
\begin{eqnarray}
\dot{\epsilon}_{T}(t) &=& -\frac{ig}{\sqrt{N_c}} \sum_{k} e^{-ikn} c_{k}(t), \label{ET} \\
\dot{\epsilon}_{C}(t) &=& -\frac{i}{\sqrt{N_c}} \sum_{k} g_{k} c_{k}(t), \label{EC} \\
\dot{c}_{k}(t) &=& -i \omega_{k} c_{k}(t) - \frac{ig}{\sqrt{N_c}} e^{ikn} \epsilon_{T}(t) - \frac{i}{\sqrt{N_c}} g_{k}^{*} \epsilon_{C}(t). \label{CK}
\end{eqnarray}
Integrating Eq.~(\ref{CK}) with the initial condition \(c_{k}(0) = 0\) (i.e., the waveguide is initially in the vacuum state), we find
\begin{eqnarray}
c_{k}(t) &=& -\frac{i}{\sqrt{N_c}} \int_{0}^{t} e^{i\omega_{k}(\tau-t)} \left( g e^{ikn} \epsilon_{T}(\tau) + g_{k}^{*} \epsilon_{C}(\tau) \right) d\tau \nonumber \\
&=& -\frac{i}{\sqrt{N_c}} \int_{0}^{t} e^{iv_{g}(|k| - \frac{\pi}{2})(\tau-t)} \left( g e^{ikn} \epsilon_{T}(\tau) + g_{k}^{*} \epsilon_{C}(\tau) \right) d\tau. \label{22}
\end{eqnarray}
Here, the dispersion relation \(\omega_k\) is linearized near \(k = \pm \pi/2\) as \(\omega_k \approx v_g (|k| - \pi/2)\), where \(v_g = 1/(2J)\) is the group velocity.

Substituting Eq.~(\ref{22}) into Eqs.~(\ref{ET}) and (\ref{EC}), we obtain
\begin{eqnarray}
\dot{\epsilon}_{T}(t) &=& -\frac{g}{N_c} \int_{0}^{t} e^{-i\frac{\pi}{2} v_{g} (\tau-t)} \sum_{k} \left( g \epsilon_{T}(\tau) + e^{-ikn} g_{k}^{*} \epsilon_{C}(\tau) \right) e^{iv_{g} |k| (\tau-t)} d\tau, \label{5} \\
\dot{\epsilon}_{C}(t) &=& -\frac{1}{N_c} \int_{0}^{t} e^{-i\frac{\pi}{2} v_{g} (\tau-t)} \sum_{k} g_{k} \left( g e^{ikn} \epsilon_{T}(\tau) + g_{k}^{*} \epsilon_{C}(t) \right) e^{iv_{g} |k| (\tau-t)} d\tau. \label{6}
\end{eqnarray}

Using the formula
\begin{eqnarray}
\frac{1}{N_c} \sum_{k} e^{iv_{g} |k| (\tau-t)} &=& \frac{2}{v_{g}} \delta(\tau-t), \label{7}
\end{eqnarray}
we simplify the dynamical equations to
\begin{eqnarray}
\dot{\epsilon}_{T}(t) &=& -\frac{g^2}{v_{g}} \epsilon_{T}(t) - \frac{gG_{2}}{v_{g}} e^{i\frac{\pi}{2}(N-n)} \Theta(t - \frac{N-n}{v_{g}}) \epsilon_{C}(t - \frac{N-n}{v_{g}}) \nonumber \\
& & - \frac{gG_{1}}{v_{g}} e^{i\frac{\pi}{2}n} \Theta(t - \frac{n}{v_{g}}) \epsilon_{C}(t - \frac{n}{v_{g}}), \\
\dot{\epsilon}_{C}(t) &=& -\frac{G_{1}^{2} + G_{2}^{2}}{v_{g}} \epsilon_{C}(t) - \frac{2G_{1}G_{2}}{v_{g}} e^{i\frac{\pi}{2}N} \Theta(t - \frac{N}{v_{g}}) \epsilon_{C}(t - \frac{N}{v_{g}}) \nonumber \\
& & - \frac{gG_{1}}{v_{g}} e^{i\frac{\pi}{2}n} \Theta(t - \frac{n}{v_{g}}) \epsilon_{T}(t - \frac{n}{v_{g}}) - \frac{gG_{2}}{v_{g}} e^{i\frac{\pi}{2}(N-n)} \Theta(t - \frac{N-n}{v_{g}}) \epsilon_{T}(t - \frac{N-n}{v_{g}}).
\end{eqnarray}

In the regime where the evolution time is much larger than the retardation time, retardation effects can be neglected. The dynamical equations reduce to \(d\vec{\epsilon}(t)/dt = -M \vec{\epsilon}(t)/v_g\), where \(\vec{\epsilon}(t) = (\epsilon_T(t), \epsilon_C(t))^T\). The coupling matrix \(M\) is
\begin{equation}
M = \begin{pmatrix}
g^2 & g \left( G_1 e^{i\phi_L} + G_2 e^{i\phi_R} \right) \\
g \left( G_1 e^{i\phi_L} + G_2 e^{i\phi_R} \right) & G_1^2 + G_2^2 + 2 G_1 G_2 e^{i(\phi_L + \phi_R)}
\end{pmatrix}
\end{equation}
where \(\phi_L = K \Delta L\) and \(\phi_R = K \Delta R\) represent the accumulated phases during photon propagation.

In Fig.~\ref{SM3}(a) and (b), we plot the eigenvalues \(\lambda_\pm\) of the matrix \(M\) for the case where the CA consists of small and giant atoms, respectively. For the small atom setup, we find that \(\lambda_- = 0\) when \(\phi_R = 2\pi\) in Fig.~\ref{SM3}(a), indicating the presence of a BIC~\cite{SL2021}. In fact, the BIC always exists as long as \(\phi_R = m\pi\) with \(m = 0, 1, 2, \dots\). In Fig.~\ref{SM3}(c), we numerically plot the photonic distribution of the BIC based on the atom-waveguide coupled Hamiltonian. The results show that the photons are localized in the 1st and 3rd resonators, consistent with the long-time behavior illustrated in Fig.~\ref{dynamics}(c) in the main text.

Furthermore, when the CA is implemented as a giant atom, we analyze the eigenvalues of \(M\) as a function of \(\phi_R\) in Fig.~\ref{SM3}(b), under the condition \(\phi_L = \pi\). The results indicate that the system also supports a BIC, evidenced by \(\lambda_- = 0\) when \(\phi_R = \pi\). The photonic distribution corresponding to this BIC is depicted in Fig.~\ref{SM3}(d), which agrees with the photonic dynamical behavior when the evolution time is long enough, as shown in Fig.~\ref{SM2}(a). However, in this case, we cannot achieve chiral radiance.

When the TA is initially prepared in the excited state, while the CA is in the ground state and the waveguide is in the vacuum state, the analytical solutions for the atomic amplitudes are obtained as
\begin{eqnarray}
\epsilon_T(t) &=& e^{-\frac{x_+ t}{2}} \left[ \cosh\left( \frac{x_1 t}{2} \right) - \frac{x_-}{x_1} \sinh\left( \frac{x_1 t}{2} \right) \right],\label{ST} \\
\epsilon_C(t) &=& -\frac{2M_{12}}{x_1} e^{-\frac{x_+ t}{2}} \sinh\left( \frac{x_1 t}{2} \right),\label{SC}
\end{eqnarray}
where \(x_+ = M_{11} + M_{22}\), \(x_- = M_{11} - M_{22}\), \(x_1 = \sqrt{x_-^2 + 4|M_{12}|^2}\), and \(M_{ij}\) are the elements of the matrix \(M\).

\begin{figure}
  \includegraphics[width=0.48\columnwidth]{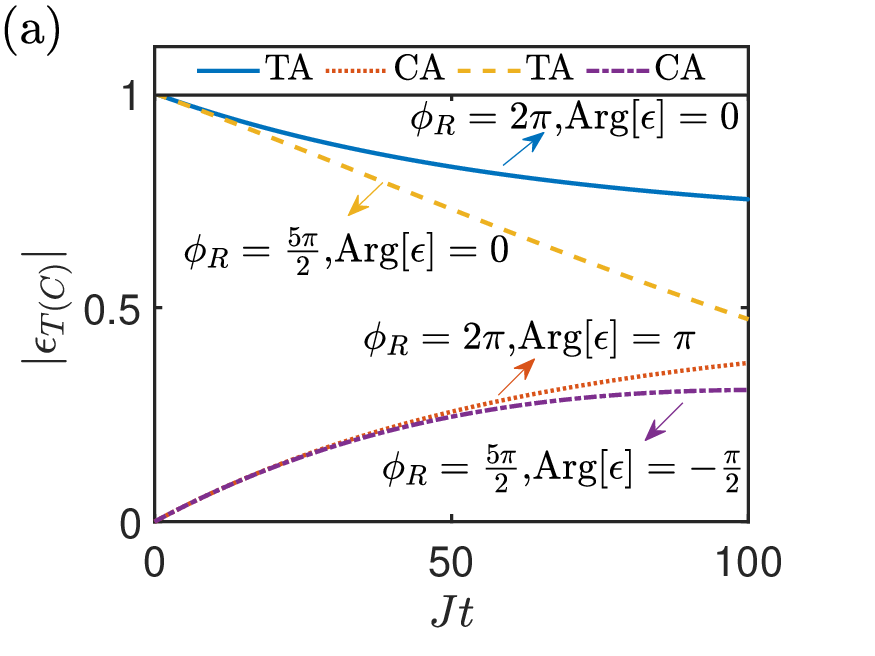}
  \includegraphics[width=0.48\columnwidth]{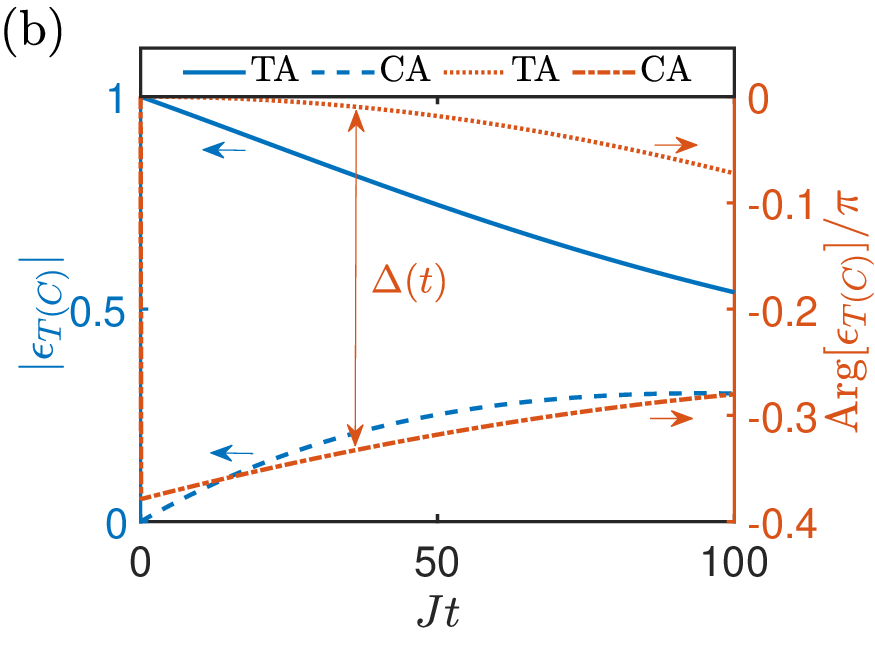}\\
  \caption{ The evolution of the atomic amplitude for small and giant CA, respectively. The parameters are set as \(\omega_T = \omega_C = \omega\), \(g = 0.1J\), \(G_{2}=0.15J\), \(N_{T}=1\), \(N_{C}=1\). (a) \(G_{1}=0\) and (b) \(G_{1}=0.067J\), \(\Delta L=2\) and \(\Delta R=5\).}
  \label{SM4}
\end{figure}

In Fig.~\ref{SM4}, we plot the time evolution of the atomic amplitude for small and giant CAs, i.e., the modulus $|\epsilon_{T(C)}|$ and the corresponding phase $\text{Arg}(\epsilon_{T(C)})$. For the small  CA setup, the phases are time dependent but this is not for the giant CA case. Furthermore, we observe that
$|\epsilon_{C}|\ll|\epsilon_{T}|$ for short time evolution. This relation will help us to discuss the chirality of the emitted photons in what follows.

Finally, we perform the inverse Fourier transform on both sides of Eq.~(\ref{22}) to derive the distribution of photons in real space. This yields
\begin{align}
c_{m}(t) &= -i \int_{0}^{t} e^{-i\frac{\pi}{2}v_{g}(\tau-t)} \sum_{k} \frac{1}{N_{c}} \left( g e^{ik(n-m)} \epsilon_{T}(\tau) + g_{k}^{*} e^{-ikm} \epsilon_{C}(\tau) \right) e^{iv_{g}|k|(\tau-t)} d\tau \nonumber \\
&= -\frac{i}{v_{g}} \Big\{ g e^{i\frac{\pi}{2}|m-n|} \Theta\left(t - \frac{|m-n|}{v_{g}}\right) \epsilon_{T}\left(t - \frac{|m-n|}{v_{g}}\right) \nonumber \\
&\quad + G_{1} e^{i\frac{\pi}{2}|m|} \Theta\left(t - \frac{|m|}{v_{g}}\right) \epsilon_{C}\left(t - \frac{|m|}{v_{g}}\right) \nonumber \\
&\quad + G_{2} e^{i\frac{\pi}{2}|m-N|} \Theta\left(t - \frac{|m-N|}{v_{g}}\right) \epsilon_{C}\left(t - \frac{|m-N|}{v_{g}}\right) \Big\}, \label{11}
\end{align}
which corresponds to Eq.~(8) in the main text.

This expression provides the photon distribution at site \(m\) as a function of time \(t\). It explicitly shows how the photon dynamics depend on the coupling constants \(g\), \(G_{1}\), and \(G_{2}\), as well as the spatial relationships between the TA, CA, and the photon at position \(m\). The step functions \(\Theta\) account for the causal propagation of photons within the waveguide.

By further neglecting the retardation effect, the photonic dynamics simplifies to
\begin{align}
c_{m}(t)
\approx -\frac{i}{v_{g}} \Big\{ g e^{i\frac{\pi}{2}|m-n|} \epsilon_{T}(t)
+ G_{1} e^{i\frac{\pi}{2}|m|} \epsilon_{C}(t)
+ G_{2} e^{i\frac{\pi}{2}|m-N|} \epsilon_{C}(t) \Big\},
\end{align}
indicating that the photonic distribution is highly sensitive to the phase of the atomic excitations.

\begin{figure}
\includegraphics[width=0.8\columnwidth]{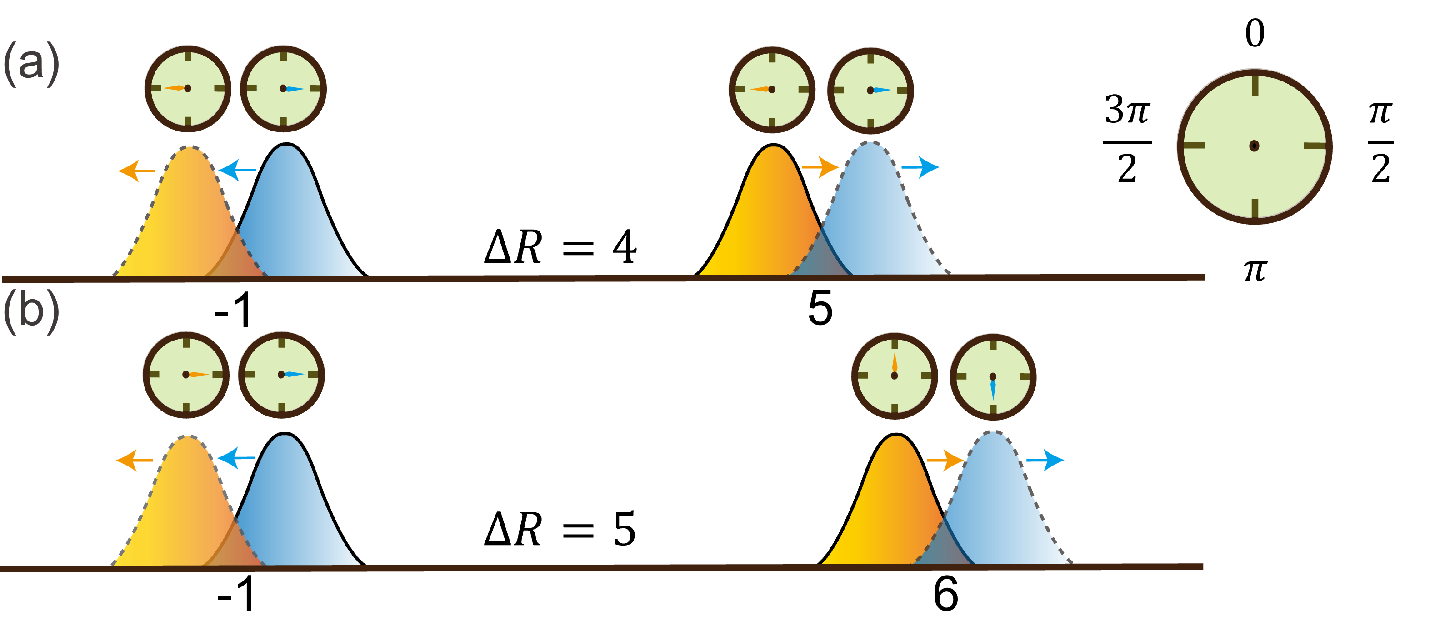}
  \caption{ Cartoon schematic diagram of photon interference for the small CA setup with (a) for $\Delta R=4$ and (b) for $\Delta R=5$. The blue (yellow) wave packet represents the wave packet generated at the TA (CA). The arrow indicates the propagation direction, and the clock reading represents the carried phase.}
  \label{chiral}
\end{figure}

This result highlights how the spatial profile of the photon field depends on the relative phase contributions from the TA and CA, mediated by their respective coupling strengths and positions within the system. For the small CA setup, we will have
\begin{eqnarray}
c_{-1}(t)&=&\frac{1}{v_{g}}(g\epsilon_{T}(t)+G_{2}\epsilon_{C}(t)),\nonumber \\
c_{N+1}(t)&=&\frac{1}{v_{g}}(g\epsilon_{T}(t)+G_{2}\epsilon_{C}(t)),
\end{eqnarray}
for $\Delta R=4$ and
\begin{eqnarray}
c_{-1}(t)&=&\frac{1}{v_{g}}(g\epsilon_{T}(t)+iG_{2}\epsilon_{C}(t)),\nonumber \\
c_{N+1}(t)&=&\frac{1}{v_{g}}(ig\epsilon_{C}(t)+G_{2}\epsilon_{C}(t)),
\end{eqnarray}
for $\Delta R=5$. Combining with the phases of the atomic amplitudes, the interference effect is demonstrated in Fig.~\ref{chiral}. For \(\Delta R = 4\), as shown in Fig.~\ref{chiral}(a), the phases of the two terms in both $c_{-1}$ and $c_{N+1}$ differ by \(\pi\), leading to destructive interference, which prevents the photon from escaping the atomic regime, consistent with the BIC physics (see SM), and we cannot observe the chirality in Fig.~\ref{dynamics}(c) of the main text. In contrast, when \(\Delta R = 5\), as shown in Fig.~\ref{chiral}(b), the two terms in $c_{-1}\,(c_{N+1})$ interfere constructively (destructively) with a \(2\pi\,( \pi )\) phase difference. As a result, the photonic intensity on the left side of the atoms is much stronger than on the right side, leading to chiral superradiance, as shown in Fig.~\ref{dynamics}(d) of the main text.

For the giant CA case of $N=7,n=2$,  that is, the CA couples to the $0$th and $7$th sites of the waveguide, while the TA couples to the $2$th sites, we will have
\begin{eqnarray}
c_{-1}(t)&=&-\frac{i}{v_{g}}\big[-g\epsilon_{T}(t)+(G_{1}+iG_{2})\epsilon_{C}(t)\big],\\
c_{N+1}(t)&=&-\frac{i}{v_{g}}\big[-ig\epsilon_{T}(t)+(iG_{1}+G_{2})\epsilon_{C}(t)\big].
\end{eqnarray}

As a result, the degree of chirality is obtained as
\begin{eqnarray}
{\eta}&=&\frac{|c_{-1}|^2-|c_{N+1}|^2}{|c_{-1}|^2+|c_{N+1}|^2}\nonumber \\
&=&\frac{2gG_{2}|\epsilon_{C}||\epsilon_{T}|\sin
[\arg(\epsilon_{T})-\arg(\epsilon_{C})]}{g^2|\epsilon_{T}|^2+(G_{1}^{2}+G_{2}^{2})|\epsilon_{C}|^2-2gG_{1}|
\epsilon_{C}||\epsilon_{T}|\cos[\arg(\epsilon_{T})-\arg(\epsilon_{C})]}
\end{eqnarray}
which yields Eq.~(\ref{etat}) in the main text by considering $|\epsilon_C|/|\epsilon_T|\ll 1$, that is,
\begin{equation}
\eta \approx \frac{2G_2 \sin \Delta |\epsilon_C|}{g |\epsilon_T|}.
\label{etatsm}
\end{equation}

\end{document}